\documentstyle[12pt,aasms4,graphics]{article}

\newcommand{\ubvri}{\protect\hbox{$U\!BV\!RI$} }
\newcommand{\bvri}{\protect\hbox{$BV\!RI$} }
\begin{document}
\oddsidemargin=0mm

\title{
The Calibration of the {\it Swift}/UVOT Optical Observations: A Recipe
for Photometry}

\author{Weidong Li, Saurabh Jha, Alexei V. Filippenko, Joshua S. Bloom, David Pooley,
Ryan J. Foley, \& Daniel A. Perley}

\affil{Department of Astronomy, University of California, Berkeley,
CA 94720-3411.\\
email: (wli,sjha,alex,jbloom,dave,rfoley,dperley)@astro.berkeley.edu}

\slugcomment{Submitted to PASP}

\begin{abstract}

The Ultraviolet-Optical Telescope (UVOT) onboard {\it Swift} has the
capability to provide critical
insight into the physics of the early afterglows of gamma-ray
bursts (GRBs). But without precise calibration of the UVOT to
standard photometric systems, it is impossible to
leverage late-time, ground-based follow-up data to the
early-time UVOT observations.
In this paper, we present a calibration of the {\it Swift}/UVOT photometry
to the standard Johnson $UBV$ system for the UVOT $U$, $B$, and
$V$ filters, and a step-by-step photometry recipe for analyzing these
data. We base our analysis on aperture photometry
performed on the ground-based and UVOT observations of the local standard stars 
in the fields of supernovae (SNe) 2005am and 2005cf, and a
number of Landolt standard stars. We find that the optimal photometry
aperture radius for UVOT data is small (2$\arcsec$.5 for unbinned data,
3$\arcsec$.0 for $2\times2$ binned data), and show that the 
coincidence-loss (C-loss) correction is important even for relatively
faint magnitudes (mag 16 to 19). Based on a theoretically motivated
model, we fit the C-loss
correction with two parameters, the photometric zero point (ZP) and 
the saturation magnitude ($m_\infty$), and derive tight constraints for 
both parameters [$\sigma(ZP) = 0.01$ mag and $\sigma(m_\infty) = 0.02$ mag)]. 
We find that the color term correction is not necessary for the UVOT
$B$ and $V$ filters, but is necessary for the $U$ filter for blue 
objects [$(U - V) < 0.4 $ mag]. 
We analyze the UVOT $UBV$ photometry of SN 2005am, and find
that the UVOT photometry is generally consistent with the
ground-based observations, but a difference of up to 0.5 mag
is found when the SN became faint. We also apply our calibration
results to the UVOT observations of GRB 050603. 
There is a scatter of $\sim$0.04--0.08 mag in our final UVOT photometry,
the cause of which is unclear, but may be partly 
due to the spatial variation in the pixel sensitivity of the 
UVOT detector. 

\end{abstract}


\section{Introduction}

The successful launch and operation of {\it Swift} heralds a new era
for the study of gamma-ray bursts (GRBs) and related phenomena. {\it Swift}, 
a multi-wavelength
space observatory, has three instruments: the Burst Alert Telescope
(BAT; Barthelmy et~al. 2005), the X-ray Telescope (XRT; Burrows et~al. 2005),
and the UV/Optical Telescope (UVOT; Roming et~al. 2005).
Together these instruments observe GRBs and their afterglows in the
gamma-ray, X-ray, and ultraviolet/optical wavebands, respectively.

Compared to previous space missions dedicated to the study of GRBs, 
the UVOT is unique to {\it Swift}, although it is identical 
to the Optical Monitor on {\it XMM-Newton} (Mason et al. 2001). The
UVOT is a 30~cm Ritchey-Chr\'{e}tien reflector,  
using micro-channel intensified CCDs as detectors. 
These are photon-counting devices capable of detecting very low signal
levels. The UVOT is designed to rapidly respond to localizations of GRBs by the 
BAT and XRT instruments. It has UV capability which is not 
possible from the ground, and it is also more sensitive than
most ground-based rapid-response telescopes.  From the UVOT images,
optical afterglows of GRBs can be quickly identified and studied,
which helps to optimize ground-based observations and provides 
information on the early-time photometric evolution of these GRB afterglows. 

It is expected that the early UVOT observations will be used in conjunction
with subsequent ground-based images.
It is thus essential that the UVOT and the 
ground-based images are calibrated on the same photometric system.
In the optical bands, the most frequently used ground-based photometric
system is the Johnson/Cousins \ubvri system, and the {\it Swift} calibration
database (CALDB) webpage \footnote{http://heasarc.gsfc.nasa.gov/docs/heasarc/caldb/swift}
provides calibration files for the various UVOT filters. Table 1 lists
these calibration results from the latest release (2005 Aug.\ 12). 
We note that the uncertainties for the zero points are relatively large
(0.1 -- 0.2 mag) for most filters. 

Since it is important to tie the UVOT photometry with that obtained
from the ground, in this paper we present an independent study of the 
photometric calibrations for the UVOT filters, in particular in the $U$,
$B$, and $V$ bands.  This calibration is derived from observations of
two supernovae, SN 2005am and SN 2005cf (but only the light curve
of SN 2005am is presented in this paper), and Landolt (1992) standard stars
in the {\it Swift} quicklook database.  The other goal of this paper is
to analyze the UVOT 
observations with tools that are familiar to optical astronomers, such
as IRAF and DoPhot, and search for optimal parameters for doing proper
photometry. The NASA High Energy Astrophysics Science
Archive Research Center (HEASARC)\footnote{http://heasarc.gsfc.nasa.gov/ .}
has supplied software tools to analyze data from all {\it Swift}
instruments. For UVOT images, these tools suggest aperture photometry
using Sextractor (Bertin \& Arnouts 1996).

The ground-based observations and reduction of SN 2005am
are described in \S 2, and photometric calibration analyses are presented in 
\S 3. Section \S 4 discusses the UVOT photometry of SN 2005am and compares
this to the ground-based observations. We apply our calibration results
and optimal photometric parameters to the UVOT observations of GRB 050603
in \S 5. The discussions are presented in \S 6 and the conclusions are
summarized in \S 7.

\section {Analysis of the Ground-based Observations of SN 2005am}

SN 2005am was discovered by R. Martin (Martin, Yamaoka, \& Itagaki 2005)
on 2005 Feb.\ 22 (UT dates are used throughout this paper) during the 
course of the Perth Automated SN Search. It was classified as a Type 
Ia SN (SN~Ia) by Modjaz et al. (2005a) from a spectrum taken with the F. L.
Whipple Observatory 1.5-m telescope. After some delays caused by bad
weather, we began to follow the SN with the robotic 0.76-m  
Katzman Automatic Imaging Telescope (KAIT; see Li et al. 2000; 
Filippenko et al. 2001; Filippenko 2003) at Lick Observatory on Mar.\ 6. Several 
epochs of observations were also obtained with the 1-m Nickel
telescope at Lick Observatory. 

Photometric calibrations of the SN 2005am field were performed 
under photometric conditions on 4 nights: Mar.\ 9 and 13 with KAIT,
and Mar.\ 12 and 14 with the Nickel telescope. During each photometric night,
many Landolt (1992) standard-star sequences (9--12 for Nickel,
16--18 for KAIT) were observed at a range of airmasses. Instrumental
magnitudes for the
standard stars were measured using aperture photometry with the
IRAF\footnote{IRAF (Image Reduction and Analysis Facility) is distributed by
the National Optical Astronomy Observatories, which are operated by the
Association of Universities for Research in Astronomy, Inc., under cooperative
agreement with the National Science Foundation.}  DAOPHOT package (Stetson
1987) and then employed to determine transformation coefficients to the standard
Johnson-Cousins \bvri system. The derived transformation coefficients and
color terms were then used to calibrate a sequence of local standard
stars in the field of SN 2005am (hereafter SN 2005am field stars). Figure 1 
shows a finder chart for 
the SN 2005am field, while Table 2 lists the magnitudes of the local standard stars and 
the root-mean-square (RMS) of the magnitude measurements in all the photometric nights.
Notice that the local standard stars have different
numbers of calibrations because the two telescopes have different total
fields of view.  The majority of the calibrated magnitudes have uncertainties
smaller than 0.03 mag. 

Also listed in Table 2 are preliminary $U$-band calibrations for some 
of the bright stars in the SN 2005am field. This calibration was done
on Apr.\ 6 under photometric conditions with KAIT, but only one \ubvri
sequence of the Landolt field
``Rubin 152" was observed at the same airmass as when SN 2005am was imaged. 
Inspection of the data also reveals that the measured 
\bvri magnitudes for the SN 2005am field stars 
from this particular night are offset from the calibration listed in 
Table 2 by a constant 0.20$\pm$0.01 mag. Further investigation suggests that
the dome was slightly blocking the telescope when the standard-star
field was imaged due to a dome position zero-point error. We thus
shift the calibrated $U$-band magnitudes by the same amount. 
We set an uncertainty of 0.05 mag to the calibrated $U$-band magnitudes,
but because we had only one standard-star sequence, and we had to apply
an arbitrary shift to the measured magnitudes, the real uncertainty could be
as high as 0.10 mag. As a result, we caution that the results derived
from the $U$-band calibration of the SN 2005am field
should be regarded as preliminary. In section \S~3.3,
we present a better $U$-band calibration for the field of 
SN 2005cf. Those data, combined with $U$-band observations of Landolt
standard stars (also in \S~3.3), lead to the final
$U$-band calibration (\S~3.4). 

As can be seen in Figure 1, SN 2005am occurred in the outskirts of its
host galaxy, and is separated from a relatively bright foreground star
by only 7$\arcsec$.  To derive proper photometry for SN 2005am, 
we use the point-spread-function (PSF) fitting method (Stetson 1987) in
the IRAF/DAOPHOT package to perform differential photometry of SN
2005am relative to the local standard stars; see Li et al. (2003a)
for more details. Color terms for the KAIT and the Nickel filters
have been well established from photometric calibrations of over 20 
photometric nights at each telescope, and have been applied to derive
the final photometry for SN 2005am as listed in Table 3.  The quoted
uncertainty of the magnitudes is a quadrature sum of the PSF-fitting
photometry and the transformation scatter from the local standard
stars. Although SN 2005am has a complex background, the final photometry
has an overall uncertainty of only 0.03--0.04 mag because the SN
is significantly brighter than the background, and there are plenty
of bright isolated stars in the field from which to construct a robust PSF
for the images.

The derived light curve of SN 2005am is shown in Figure 2, together with 
fits using the Multicolor Light Curve Shape (MLCS2k2) method (Jha 2002; Jha, Riess, \&Kirshner 2006),
which is an empirical method to model the light curves of a SN~Ia
to derive its luminosity distance. Overall, the KAIT and the Nickel
photometry are consistent with each other (except perhaps the two 
$I$-band data points near JD 2453470). SN 2005am shows 
a photometric evolution rather typical of a SN~Ia: a second peak
in the $I$ band, a shoulder in the $R$ band, and a smooth decline
after maximum in the bluer bands. Our follow-up observations began near the 
maximum of the $B$ band, and 2--3 days before maximum in 
the other bands. 
The MLCS2k2 fits are typical for a well-observed SN~Ia. SN 2005am is not
significantly reddened by dust in its host galaxy (host $A_V = 0.09 \pm
0.07$ mag). It is also a somewhat rapidly declining and subluminous object
(by approximately 0.5 mag), intermediate between normal SNe~Ia and the most
subluminous objects like SN 1991bg (Filippenko et al. 1992). This makes SN 2005am an important
addition to the sample of nearby SNe~Ia, with only a handful of similar
objects known (Jha et al.~2005).

The SN 2005am field stars as listed in
Table 2 are used to study the photometric calibrations
of the UVOT filters in \S 3. We investigate the optimal 
parameters to do photometry on these stars in the UVOT images, 
so that the best possible consistency between the ground-based 
KAIT and Nickel calibrations (hereafter the ``Lick calibration'') and
the UVOT measurements can be achieved. 
These stars cover a wide range of brightness (from $B$ = 12.30
to $B$ = 19.03 mag) and color [from $(B - V)$ = 0.51 to
$(B - V)$ = 1.39 mag]. The photometry of SN 2005am itself will
provide ground-based estimates for the magnitudes of SN 2005am 
at the epochs of the UVOT observations, as discussed
in \S 4.

Since the filter transmission curve is a good indication of how 
standard a filter is, in Figure 3 we show a comparison of the
transmission curves for the 
$U$, $B$, and $V$ filters involved in our analysis, including these
used by KAIT (thin solid lines), Nickel (dash-dotted lines), and
UVOT (dashed lines). Also plotted are the standard Johnson-Cousins
$UBV$ transmission curves (thick solid lines) as described by Bessell (1990). 
For the $B$ and $V$ filters, the transmission curves generally share the
same spectral range and are similar to each other. In particular,
the UVOT $B$ and $V$ filters are quite consistent with the
Bessell descriptions, so we expect the color terms for
these filters to be small, as confirmed later in the paper.
For the $U$ filter, however, the one used by KAIT is quite different
from the one used by UVOT, and their transmission curves are
quite different from the Bessell description. Relatively large color
terms for these filters are thus expected, as discussed below. 

\section{Photometric Calibration of the UVOT}

\subsection{UVOT Observations of SN 2005am}

A journal of UVOT observations of SN 2005am is listed in 
Table 4. These are the data available to the general users
after {\it Swift} data were made public on 2005 Apr.\ 1.
A more complete set of UVOT data on SN 2005am, some of which were not 
made available to us, is reported
by Brown et al. (2005a).
We first retrieved the data from 
the {\it Swift} quicklook database, and in all cases, the level-2 filter sky images
were downloaded. From the {\it Swift} manual, the level-2 data are 
what most researchers will use to start their analysis. The UVOT reduction
pipeline has been performed on these images, which are also 
stored in sky coordinates (RA$_{\rm J2000}$ and DEC$_{\rm J2000}$). 
The accompanying exposure maps for each individual image are also
 downloaded. 
There are UVOT observations of SN 2005am in other filters
($UVW1$, $UVW2$, and $UVM2$) as well, but these images are not
listed in Table 4, and will not be studied in this paper since
these filters are in the far-UV where we do not have ground-based
calibrations.

There are five $UBV$ sequences observed by UVOT in Table 4, which we hereafter
refer to as 
obs1, obs2, obs3, obs4, and obs5, respectively. The $U$-band 
observation in sequence obs1 will be referred as ``obs1 $U$," etc. 
There is only a short (18.02~s) $U$-band exposure, and a normal $V$-band
exposure in obs4;
the $B$ band is missing.
All sequences were observed without binning except obs2, in which 
a $2\times2$ on-board binning was used. The SN was well detected in 
all images except obs4 $U$ and obs5, for which the exposure times
were too short for the brightness of the SN. The unbinned UVOT data
have a resolution of 0.5$\arcsec$ per pixel, and the total field
of view is 16$\arcmin$.4$\times$16$\arcmin$.4. Since the fields of view
of KAIT (6$\arcmin$.6$\times$6$\arcmin$.6) and the Nickel telescope
(6$\arcmin$.3$\times$6$\arcmin$.3) are much smaller, the field for which
we have ground-based calibration is only a fraction of the total 
UVOT field. 

Inspection of the UVOT images reveals two things worth
noting for observers familiar with the reduction of ground-based CCD observations.
First, as shown in Figure 4 and also indicated in the {\it Swift} manuals,
the PSFs of the stars vary with the count
rate (i.e., magnitude) of the object and with the filter being used, and they
may vary with position
on the detector. The brightest stars also show various
degrees of ``ghost" emission, including ghost wings, ghost rings, 
and rings around the stars themselves. This is very different from
ground-based CCD images, in which stars of different brightness 
have a constant PSF across the image (except for image
distortion in wide-field images, optical defects, or stars
that are saturated). As discussed more in later sections,
a varying PSF is a serious challenge for doing
stellar photometry, unless the intrinsic PSFs can be constructed 
according to the brightness of the stars and their positions on 
the detector.  Unfortunately, the information on the intrinsic
PSF was not available at the time when we conducted this study.

Another aspect of the UVOT images that is different from
ground-based CCD observations is the sky background distribution and
the associated noise pattern.  Figure 5 shows the histograms
of the sky background distribution around star \#5 in the first four 
$U$-band observations. The histogram in obs5 $U$ is not shown, 
but it is similar to that of obs3 $U$. These background values are extracted 
from an annular region with an inner radius of 35 pixels and 
an outer radius of 45 pixels around star \#5 for unbinned data,
and an inner radius of 17.5 pixels and an outer radius of
22.5 pixels for the $2\times2$ binned obs2, as discussed more
in the next section. As shown in the figure, only in obs2 does 
the sky background around star \#5 show a Gaussian distribution.
In the other three observations, due to the low background,
the histogram peaks and truncates at background value 0, and
shows a Poissonian rather than a Gaussian distribution.  This is
different from ground-based CCD observations in which the
background distribution closely follows photon 
statistics and is mostly Gaussian. How to optimally account 
for the background contamination at the location of stars has
direct impact on the photometry, as discussed more in the 
next section.

\subsection{Aperture Photometry Parameters}

In this paper we use $U$, $B$, and $V$ as the magnitudes in the
standard Johnson/Cousins \ubvri system, such as the calibrated
magnitudes for the local standard stars listed in Table 2. 
We use $u$, $b$, and $v$ as the instrumental magnitudes 
measured from the UVOT image.  The ``phot" task in the
IRAF/DAOPHOT package is used to
carry out the aperture photometry throughout the paper. 
The parameters derived in our paper should be easily 
adaptable to other photometry programs that work on FITS images. 

In theory, if one can properly estimate the sky background and use
a very large aperture to sum all the flux, a varying PSF
should not be a problem for doing photometry.  In practice, however,
a big aperture includes more sky background and its associated
noise.  The signal-to-noise ratio (S/N) for the photometry will thus diminish,
so large apertures only work for bright objects. Moreover, because of contamination
from neighboring objects, it is often not possible
to use large sky regions or apertures when doing photometry.

For the reductions used throughout this paper, we adopt an annular sky 
background region with an inner radius of 35 pixels
and an outer radius of 45 pixels centered on each object for unbinned data, and 
an inner radius of 17.5 pixels and an outer radius of 22.5 pixels for
$2\times2$ binned data. The full width at half maximum (FWHM) of
the UVOT images is about 4.0 pixels (2.5 pixels for the $2\times2$
binned data), so the sky region starts more than (7--8) $\times$ FWHM
from the source, which is further than one generally uses for ground-based 
photometry [(4--5) $\times$ FWHM]. We thus expect the contamination
from the source itself to the background to be small. For bright
objects that have ghost emission as shown in Figure 4, there is 
considerable emission from the source itself in our defined sky
region, and we will attempt to evaluate how our analyses can be 
applied to bright objects in \S 3.4. Fortunately, for all the local
standard stars in the field of SN 2005am, only star \#6 is bright
and shows a surrounding ring in most of the images. We 
thus exclude star \#6 in our studies in the following sections,
but include it in the studies for bright objects in \S~3.4.

\subsubsection{Optimal Aperture Size}

What aperture size should one use in the 
``phot" program, so that the measured instrumental magnitudes 
are most consistent with the Lick calibration? To answer this question, we
perform photometry for the SN 2005am field stars
using aperture radii of 1 to 25 pixels. 
As a starting point, we use the zero points for the UVOT
filters from Table 1. For each aperture radius, the difference between the 
UVOT photometry and the Lick calibration is calculated
for each local standard star, the average difference is 
calculated, and the RMS around this average is determined.
Since the average difference can be corrected by changing
the zero point, the RMS around the average difference
measures the degree to which the UVOT photometry is consistent
with the Lick calibration.

The DAOPHOT package offers various sky background fitting algorithms;
see Stetson (1987) for detailed discussions. In this section, we use a 
simple ``mean" method to determine the sky background.
\S 3.2.4 explores the other algorithms, and concludes 
that for UVOT observations of SN 2005am,  ``mean" works best for
stars that are not close to other contaminating sources. 

Figure 6 shows the RMS versus the aperture size (radius) in 
pixels. The open circles are for obs1, the solid circles for 
obs2, the stars for obs3, the solid triangles for obs4, and the
solid line for obs5. For obs2
(binned $2\times2$), the apertures shown in Figure 6 
(APT$^\prime$; solid circles) are the actual aperture sizes
(APT) scaled to match the unbinned data. We find the best
match to be APT$^\prime$ = APT$\times$2 $-$ 1, rather than
APT$^\prime$ = APT$\times$2 as one would expect. The cause 
of this difference is unclear, but we note that the FWHM of obs2
($\sim$ 2.5 pixels) is not exactly half of the other unbinned
observations (FWHM $\sim$ 4.0 pixels) either.

Figure 6 indicates that for the $B$ and $V$ bands, the smallest
RMS ($\sim$ 0.07 mag) is achieved with an aperture size of 
5 pixels (3 pixels for $2\times2$ binned data). This result is 
consistent with the trend found in ground-based CCD images,
where the best S/N is often achieved
when the photometry aperture is close to the FWHM. 
For the $U$ band, the RMS shows a rather flat distribution
for apertures in the range 5--12 pixels. 

An aperture size of 12 pixels is used in the {\it Swift} manual to 
determine the zero points. As shown in Figure 6, however, the RMS
at 12 pixels is 0.04--0.10 mag larger than at 5 pixels 
(up to 0.15 mag larger in obs3 $V$) for the 
$B$ and $V$ bands, and is about the same for both apertures 
in the $U$ band. 

We thus find that an aperture radius of 5 pixels (3 pixels for
$2\times2$ binned data) gives the most consistent results 
between the UVOT photometry and the Lick calibration of the 
SN 2005am field stars. In sky coordinates, this is 2$\arcsec$.5, and 
3$\arcsec$.0, respectively. 

\subsubsection{Preliminary Zero Points}

Ideally, the optimal aperture size also gives the most consistent
zero point (a source yielding 1 count per second) for the observations.
The average differences
as calculated in the previous section, which represent the
amount to which the UVOT zero point in Table 1 needs to be
modified when a specific aperture is used, is plotted in Figure 7. 
The same symbols are used for the different images as in Figure 6. 

We note that with the exception of obs3 $V$ (stars in the lower 
panel) and obs4 $U$ (triangles in the upper panel), 
the other curves all converge in the aperture
size range of 4--8 pixels, then diverge when the aperture is larger
or smaller. An aperture size of 5 pixels indeed gives a very
consistent zero point for these observations. 

Obs4 $U$ is a short exposure (18.02~s),
with most of the background having a value of 0; the local standard
stars are not well detected. We remove obs4 $U$ from the zero point
determination for the $U$ band, and caution that our 
$U$-band zero point may not work for short exposures. 

It is puzzling that the well-observed obs3 $V$
gives a different zero point than the other
observations. We compare this image to obs1 $V$.
Using a large aperture size of 35 pixels, we measure the total
flux for several bright (but without ghost emission) stars, and find
that the flux ratios between these stars are inconsistent with the 
ratio of the exposure times listed in Table 4. From the flux ratios
of the stars and the exposure time of obs1 $V$,
the matching exposure time for obs3 $V$ is about
68~s, rather different from 82.77~s as listed in Table 4.
The {\it Swift} UVOT team (Brown et al. 2005c) reported that for GRB 050603, 
one UVOT exposure was affected by an error in the on-board
shift-and-add code, which resulted in a large amount of 
missing data and an effective exposure time that is much less
than indicated in the FITS header. The ``UVOT Digest"
page\footnote{http://swift.gsfc.nasa.gov/docs/swift/analysis/uvot\_digest.html}
further announced that the exposure time keywords are incorrect
for a small fraction of the UVOT images, and a list of such images is 
provided\footnote{http://swift.gsfc.nasa.gov/docs/swift/analysis/unmatched\_exposures.txt .}.
Obs3 $V$ is one of the images with wrong exposure time keywords.
The corrected exposure time is 60.0~s, slightly
different from what we derived from comparing the flux ratio 
between obs1 $V$ and obs3 $V$ ($\sim$ 68~s). 

With obs3 $V$ and obs4 $U$ excluded from the analysis, we measure the
following first-order zero points for the UVOT $UBV$ filters, when an aperture
size of 5 pixels (3 pixels for the $2\times2$ binned data) is used:
$ZP(U) = 18.22\pm0.10$ mag, $ZP(B) = 18.88\pm0.09$ mag, and 
$ZP(V) = 17.67\pm0.07$ mag. 
The uncertainty of the zero point is the quadrature sum of the RMS from 
the multiple observations and the RMS of the differences between
the UVOT photometry and Lick calibration shown in Figure 6. 
These zero points will be refined in \S~3.4. We note here that these
zero point uncertainties are overestimated, since part of the RMS between the 
UVOT photometry and Lick calibration is caused by an intrinsic 0.04--0.05
mag scatter in the UVOT photometry (\S~6.2). Moreover, the UVOT photometry
has not been corrected for the coincidence-loss correction, which is 
necessary even for these relatively faint SN 2005am field stars,
as we discuss in \S~3.4. 

We also note that when an aperture size of 12 pixels is used,
the zero points for the UVOT $UBV$ filters from our analysis 
are $ZP(U) = 18.49\pm0.14$ mag, $ZP(B) = 19.16\pm0.20$ mag, and
$ZP(V) = 17.92\pm0.18$ mag. These zero points are consistent
with those from the {\it Swift} calibration database as listed in Table 1
to within the quoted errors.  Note that the uncertainties of
the zero points with an aperture size of 12 pixels is significantly
larger (more than double in $B$ and $V$) than those with an aperture
size of 5 pixels. \emph{We strongly recommend the use of 5-pixel
apertures for all UVOT UBV photometry.}

\subsubsection{The Necessity of the Coincidence-Loss Correction}

In an attempt to further refine the zero points, 
the residuals of the UVOT photometry when compared to the Lick
calibration have been vigorously studied. The UVOT photometry
is measured using the optimal aperture size and zero points as
discussed above.

In Figure 8, we show the residual of $b(UVOT) - B(Lick)$ versus the
$b(UVOT) - v(UVOT)$ color for the local standard stars in obs1 $B$. There 
is no apparent correlation between the residuals and the 
colors, suggesting that the presence of a large color term is unlikely.

In the upper panel of Figure 9, we show these residuals again, 
but as a function of $b(UVOT)$. A strong correlation 
can be seen in this plot. To account for this correlation, a correction
factor ($CF$) needs to be multiplied to the magnitudes measured in the 
``phot" program. The middle panel of Figure 9 shows 
that when $CF = 1.07$ is applied to correct for the measured UVOT
photometry (with a new zero point of 17.77 mag), the RMS of the
photometric differences is significantly improved, from
RMS = 0.082 mag to RMS = 0.035 mag. 

We analyze all the $B$-band and $V$-band images (except obs3 $V$),
and a strong dependence of the residuals on the magnitude
is apparent for all the images. The correction factors measured from
the images are consistent for the same band (to within errors), 
but are different for the two bands.
We did not attempt to measure a correction factor for the $U$ 
band because we only have a preliminary $U$-band calibration 
for a limited number of stars.

We have performed several analyses to investigate the properties 
and the possible cause(s) of the correction factor. We find that
the correction factor is related to the magnitude (or count rate for
the UVOT detectors) rather than the total counts of the final detection
(which is a function of both count rate and exposure time). 
The correction factor is present in the data reduced with different sky regions
or aperture sizes, suggesting that it is not 
caused by sky background evaluation, or by our choice of a relatively
small optimal aperture. 

We find that the correction factor is a natural consequence of the 
coincidence-loss (hereafter C-loss) correction; see the detailed 
discussion in \S~3.4. The bottom panel of Figure 9 shows the
residuals of the obs1 $B$ photometry after the C-loss correction,
and no apparent correction is present. The RMS after the C-loss
correction is 0.041 mag, comparable to that achieved using 
the correction factor (0.035 mag). This indicates that even 
for relatively faint (mag 16 to 19) stars, it is necessary
to consider the C-loss correction. Since the correction factors
and their corresponding zero points can be naturally accounted
for in the C-loss correction (\S~3.4), we do not report their 
values separately here. 

\subsubsection{Optimal Sky Fitting Algorithm}

At the low background count rates found in the UVOT, the 
distribution of the background in a given aperture is 
Poissonian, rather than Gaussian as in ground-based optical
images. Because of this, some of the usual sky-fitting routines
in IRAF are inappropriate to use. 
In this section we explore the various sky background fitting algorithms
offered by IRAF/DAOPHOT. As discussed in more detail by Stetson (1987),
DAOPHOT offers the following sky-fitting methods: constant, file,
mean, median, mode, centroid, gauss, ofilter, crosscor, histplot, and 
radplot. Among these, ``constant" and ``file" require background values
supplied by the user, and are not adopted in our reductions. ``Gauss"
fits a Gaussian function to the background histograms. As shown
in Figure 5 and as discussed in \S 3.1, most of the histograms 
do not show a Gaussian distribution, so this method is not used.
``Histplot" and ``radplot" require the user to mark the background
interactively, on the histogram and radial profile plot of the background,
respectively. We found it difficult to visually estimate a reasonable
background from the plots for many of the observations, and did not
include the results from these two methods in our analysis. 

Among the six sky-fitting algorithms we used (mean, median, mode, centroid, 
ofilter, and crosscor), ``mean" often outputs the largest 
background value, while ``centroid" outputs the smallest. The ``centroid" method
often estimates a background value of 0 (as does the ``mode" method),
since that is where the histogram peaks for many observations; thus, it
produces the brightest magnitude measurements among
all the methods. 

Using our defined sky background region and the optimal aperture size,
we analyze obs1 $V$ using all six sky-fitting algorithms.
We derive the zero points and their uncertainties,
and the correction factor for each sky-fitting method, and list
the results in Table 6. Without using $CF$, all the methods yield
consistent zero points and similar uncertainties. The magnitudes
measured with the ``centroid" method are on average 0.04 mag brighter
than with the ``mean" method, as indicated by the difference in the
zero points. When $CF$ is used, different methods require slightly
different values of $CF$ (and the associated zero points).  The uncertainties of the
zero points are very similar, and are also significantly smaller than
those without using $CF$.

The ``mean" method offers a slightly smaller uncertainty than
the other methods, and also requires the smallest $CF$; it is thus
preferred by us. We note, however, that the results from different 
sky-fitting methods do not differ significantly in terms of 
zero-point uncertainties, at least for an uncrowded field 
such as the SN 2005am field stars. 
In \S 4, we attempt to derive the photometry for SN 2005am 
in the UVOT images.  In this case the different sky-fitting methods
yield more significant differences, as SN 2005am is contaminated
by a bright nearby star and its host galaxy.

\subsection{Observations of SN 2005cf and Landolt Standard Stars}

It is important to verify that the photometric parameters we derived
from the observations of SN 2005am also work for other
UVOT observations, and to include more standard stars to improve
the calibrations. For this purpose we have included observations
of SN 2005cf, for which the UVOT has multiple-epoch observations
in the various filters, and for which we have good ground-based
follow-up observations and calibrations. 

SN 2005cf was discovered by Pugh \& Li (2005) 
during the course of the Lick Observatory Supernova Search (LOSS; Filippenko 2005)
on 2005 May 28. It was classified as a SN Ia more than 
10 days before maximum light by Modjaz et al. (2005b)
from a spectrum taken with the F. L. Whipple Observatory 1.5-m telescope
on 2005 May 31. SN 2005cf was chosen as a target for the {\it Hubble Space Telescope\/}
({\it HST}) GO program 10182 (PI: A. V. Filippenko), and was extensively followed 
by many groups using ground-based telescopes and by {\it Swift}/UVOT.   
A paper on the photometry of SN 2005cf will combine the data from {\it HST}, 
{\it Swift}/UVOT, and the ground-based telescopes (Li et al. 2006,
in prep.). 

Photometric calibrations of the SN 2005cf field were performed
under photometric conditions on 4 nights: 2005 June 3 with KAIT,
and 2005 June 3, July 8, and July 11 with the Nickel telescope. The 
same observation and reduction procedures as described for SN 2005am
in \S~2 were followed to derive the standard \ubvri photometry for the 
local standard stars in the field of SN 2005cf. Table 6 lists the 
$U$, $B$, and $V$ magnitudes of the local standard stars, and 
Figure 10 shows a finder chart for the SN 2005cf field. On the 
night of 2005 July 11 a field (the right panel in Figure 10) that
does not include SN 2005cf, but is still included in the UVOT
field of view for SN 2005cf, has also been calibrated; this is also
the only night that the two fields have been calibrated in the $U$ band. 
An arbitrary 0.03 mag error is assigned for the stars that have been
calibrated only once. Some stars in Table 6 are listed as having been
calibrated 5 times in the 4 photometric nights since the SN 2005cf
field was observed twice in the $B$ band on the night of 2005 July 8. 

There are many observations of SN 2005cf by UVOT, 
and we chose the images that have the longest exposure times 
and also have a complete $U$, $B$, and $V$ sequence in our 
analysis. These data are listed in  Table 7. 

We also searched the {\it Swift} quicklook database, found
observations for the Landolt standard-star fields ``SA104" and
``SA95," and included these in our analysis (Table 8). Many of the 
Landolt standard-star observations have multiple exposures in 
a single sequence (Column 5 of Table 8), providing multiple
measurements for the magnitudes of each individual standard star. 

For all the images listed in Tables 7 and 8, we identify
the local standard stars in the field of SN 2005cf
and the Landolt standard stars, 
and perform photometry using the optimal parameters
derived in the previous sections: the sky background is 
determined using the ``mean" method in a region that is 
35 to 45 pixels from the center, and an aperture size of 
5 pixels is used. For the photometric zero points, we initially
used $ZP(U)$ = 18.22 mag, $ZP(B)$ = 18.88 mag, and
$ZP(V)$ = 17.67 mag as derived in \S~3.2.2, but changed these
to the final zero points as derived in \S~3.4: $ZP(U)$ = 18.24 mag,
$ZP(B)$ = 18.92 mag, and $ZP(V)$ = 17.69 mag. Adopting the
final zero points enables us to directly compare the 
difference between the UVOT and the standard photometry. 

Since there are multiple measurements for the photometry of
most standard stars, we average and list their magnitudes in 
Table 10. For the Landolt standard stars, we exclude those 
that have been calibrated only once, and those that have 
calibration errors in excess of 0.05 mag in any of the 
$U$, $B$, or $V$ bands. We also include in Table 10 the 
averaged photometry for the SN 2005am field stars, but 
did not include the photometry from the sequence obs2 
(binned $2\times2$) or obs3 (exposure time error). 
Since the preliminary $U$-band calibration for the SN 2005am
field is inferior to that for the SN 2005cf field, we 
did not include the $U$-band reduction for the SN 2005am
field. The errors for the average magnitudes
as listed in Table 10 are only the RMS of the multiple 
measurements, and do not include the photometric error of 
the individual measurements: we find that when the average 
photometric error of the individual measurements is added in
quadrature to the error, the final uncertainties are overestimated,
as the reduced $\chi^2$ is less than unity for the C-loss correction
fit in the next section.  
Since most of the magnitudes have many measurements (8--49),
the RMS around the average is probably a more accurate
estimate of the true photometric uncertainty. Table 10 provides
the basis for deriving the C-loss correction and the final zero
points in the next section. 

\subsection{The C-loss Correction and the Final Photometric Zero Points}

In the upper panels of Figures 11, 12, and 13, we compare the UVOT 
to the standard photometry in the $U$, $B$, and $V$ bands, respectively, 
for all the stars listed in Table 10. The dashed lines in these panels
represent the relation where the photometry in the two systems is equal.
It can be seen that when the stars get progressively brighter, the 
UVOT photometry becomes more deviant from the standard photometry, and
reaches a limit. This is caused by the C-loss. Here we present a brief
introduction to the C-loss and how it is modeled in this paper.

The UVOT detector is a photon-counting device.  The photon counter 
integrates for a short time interval (the readout rate; 11~ms for 
UVOT); if zero photons arrive during
this period, the detector records zero counts. If one photon arrives
in this time interval, the detector records one count. {\it If more than
one photon arrives in the interval, the detector still records only 
one count.} Over a large number of integrations, for a source which sometimes
provides more than one photon per time interval, the total measured 
counts will be {\it less} than the true counts from the source. This is
the cause of the C-loss. 

The arrival of photons is a Poisson process. For a source having
a true count rate of 
$\mu$, the probability of getting $n$ photons in the time interval is 
given by 
\begin{equation}
P(n) = \frac{\mu^n}{n!}e^{-\mu} \qquad \qquad n = 0,1,2,3,\ldots
\end{equation}
Without C-loss, if we average over many time intervals, the average measured
count rate $\langle n \rangle$ would be
\begin{equation}
\langle n \rangle = 0 \cdot P(0) + 1 \cdot P(1) + 2 \cdot P(2) +
3 \cdot P(3) + \ldots = \sum_{n=0}^{\infty} nP(n) =  \sum_{n=0}^{\infty}
\frac{n\mu^n}{n!}e^{-\mu} = \mu,
\end{equation} 
which is the true count rate. 

With C-loss, however, the measured average count rate $x$ is different from Equation
2. In this case, for $n \geq 2$, the detector records a count of only 1 instead of
n, so
\begin{equation}
x = 0 \cdot P(0) + 1 \cdot P(1) + 1 \cdot P(2) + 1 \cdot P(3) + \ldots
= 0 + \sum_{n=1}^{\infty} P(n).
\end{equation}

\noindent
We note that the sum of $P(n)$ from zero to $\infty$ must be 1 (because those
are all the possibilities for $n$), so then we have

\begin{equation}
x = \sum^\infty_{n=1} P(n) =\sum^\infty_{n=0} P(n) - P(0) = 1 - e^{-\mu}. 
\end{equation}

\noindent
This means that for a photon-counting device with C-loss, such as the UVOT
detector, a source that has a true count rate of $\mu$ yields a measured
count rate of $1 - e^{-\mu}$. 

To model the C-loss, we introduce two parameters: ZP and $m_\infty$.
ZP is the photometric zero point for the UVOT data, such that a source
yielding 1 count per second has a magnitude of 0:

\begin{equation}
m_{\rm UVOT} = -2.5 \log_{10} \left( \frac {\rm counts}{\rm sec} \right) + {\rm ZP}. 
\end{equation}

\noindent
Let $m_\infty$ be the Landolt magnitude that gives a true count rate of 1. 
Since the measured count rate ($1 - e^{-\mu}$) can only reach 1 when 
$\mu$ is infinity ($\infty$), $m_\infty$ is also the UVOT
magnitude that would be measured from an infinitely bright star (the
saturation magnitude). For a star that has a Landolt magnitude of
$m_{\rm Landolt}$ and a UVOT magnitude of $m_{\rm UVOT}$, we thus have

\begin{equation}
\mu = 10^{-0.4(m_{\rm Landolt} - m_\infty)},
\end{equation}
\begin{equation}
1 - e^{-\mu} = 10^{-0.4(m_{\rm UVOT} - m_\infty)}. 
\end{equation}

\noindent
>From these two equations we can convert the magnitudes between the 
UVOT and Landolt systems: 

\begin{equation}
m_{\rm UVOT} = -2.5 \log_{10}(1-e^{-\mu}) + m_\infty, \ {\rm where}\ \ \  \mu = 10^{-0.4(m_{\rm Landolt} - m_\infty)}; 
\end{equation}

and 

\begin{equation}
m_{\rm Landolt} = -2.5 \log_{10}(\mu) + m_\infty, \  {\rm where}\ \ \  \mu = -\ln\, (1-10^{-0.4(m_{\rm UVOT}-m_\infty)}).
\end{equation}

We use Eqs. 5 and 8 to fit all the data in Table 10. A $\chi^2$-minimizing
technique is used, and the UVOT magnitudes that are too close to $m_\infty$
are not included in the fit (the open circles in Figures 11 to 13). The Landolt
standard star 95\_42, which has a very blue color ($U-V$ = $-$1.33 mag),
is not included in the $U$-band and $B$-band fits, but is used in 
the analysis of the color terms in \S~3.5. The best fits to the data are 
plotted as the solid lines in the upper panels of Figures 11 to 13, and 
the residuals around the best fits are plotted in the lower panels. The fit
parameters are also labeled in the upper panels of the figures. For the final
photometric zero points, we derive $ZP(U)$ = 18.24 $\pm$ 0.01 mag, 
$ZP(B)$ = 18.92 $\pm$ 0.01 mag, and $ZP(V)$ = 17.69 $\pm$ 0.01 mag.
For $m_\infty$, we derive $m_\infty(U)$ = 13.43 $\pm$ 0.02 mag, 
$m_\infty(B)$ = 14.16 $\pm$ 0.02 mag, and $m_\infty(V)$ = 12.92 $\pm$ 0.02 mag.
The RMS values of the fits are RMS($U$) = 0.080 mag, RMS($B$) = 0.044 mag, and
RMS($V$) = 0.045 mag, and the reduced $\chi^2$ values are $\chi^2(U)$ = 0.74,
$\chi^2(B)$ = 0.79, and $\chi^2(V)$ = 1.64, respectively. The reduced $\chi^2$
for the $V$-band fit (1.64) is a slightly bigger than unity, but is dominated
by only a few data points. For example, removing the first two points
whose UVOT photometry is close to $m_\infty$ reduced the $\chi^2$ to 1.24.

Our fits to the data in Table 10 provide tight constraints to the photometric
zero points ($\sigma$ = 0.01 mag) and $m_\infty$ ($\sigma$ = 0.02 mag).
Applying the C-loss correction also demonstrates that we can derive reliable photometry
even when the C-loss is significant (as shown by the residuals in the 
lower panels of Figures 11 to 13, and the RMS of the fits). Note 
that $m_{\rm UVOT} \leq m_\infty$
is not allowed because it corresponds to a true flux that is infinite. Stars 
that give UVOT magnitudes too close to $m_\infty$ should be treated with caution
as well. The dash-dotted lines in the lower panels of Figures 11 to 13 indicate
the Landolt magnitudes that would give a UVOT magnitude 0.1 mag bigger
than the corresponding $m_\infty$: $U$ = 12.4 mag, $B$ = 13.2 mag, and
$V$ = 12.0 mag. We tentatively consider these as the limiting bright 
magnitudes to derive reliable photometry from the UVOT data. 

We note that the RMS of the C-loss correction fits, $\sim 0.04$ mag 
for $B$ and $V$, and $\sim 0.08$ mag for $U$, represent the photometric
precision we can achieve for the UVOT images. The lower panels of Figures 11
to 13 show that although the scatter (and the corresponding error bars) 
around the fits are bigger when the stars become fainter, scatter
is present in the magnitudes for relatively bright stars as well. More 
discussion of the photometric scatter can be found in \S~6.2. 

\subsection{The UVOT Color Terms}

So far in this paper we have treated the UVOT $U$, $B$, and 
$V$ filters as if they follow exactly the standard prescription. As shown in Figure 3,
however, the transmission curves for the UVOT filters are somewhat
different from the Bessell descriptions, so some color terms
are expected. These color terms are listed in
the latest calibration data release (2005 Aug.\ 12) as follows:

\begin{equation}
 U - V = 0.087 + 0.8926 (u - v) + 0.0274 (u - v)^2,
\end{equation}
\begin{equation}
 B - V = 0.0148 + 1.0184 (b - v),
\end{equation}
\begin{equation}
 B = b + 0.0173 + 0.0187 (u - b) + 0.013 (u - b)^2 - 0.0108 (u - b)^3 - 0.0058
 (u - b)^4 + 0.0026 (u - b)^5, ~{\rm and}
\end{equation}
\begin{equation}
 V = v + 0.0006 - 0.0113 (b - v) + 0.0097 (b - v)^2 - 0.0036 (b - v)^3.
\end{equation}

\noindent
The uncertainties of the coefficients have not been reported.
These equations are only valid for the range of colors 
$-1.468 < (U - B) < 1.804$ mag, $-1.852 < (U - V) < 3.306$ mag,
and $-0.384 < (B - V) < 1.642$ mag.

For the color ranges in which these equations are valid, 
the difference between $B$ and $b$ (Eq. 12),
and also between $V$ and $v$ (Eq. 13), is $< 0.02$ mag, significantly smaller than
the photometric precision that we can achieve in our study ($\sim 0.04$ mag).
The color term for $(B - V)$ (Eq. 11) is also small. We thus 
expect the errors introduced into the UVOT photometry by treating
the $B$ and $V$ filters as standard to be small. In Figure 14 we
show the residuals
of $(B - V) - (b - v)$ versus $(b - v)$, where ($B - V$)
are the colors in the standard system, and ($b - v$)
are the UVOT photometric colors after the C-loss correction. 
Also overplotted in Figure 14 are three fitting functions: 
the solid line is $(B - V)$ = $(b - v)$, which
is used in this paper; the dashed line is
$(B - V) = 1.0148 (b - v) - 0.014$, which is adopted from
Eq. 11 but shifted up and down to best match the data; and
the dot-dashed line is our best fit to the
data: $(B - V) = 0.989 (b - v) + 0.004$.  The RMS values
for the three fits are 0.054 mag, 0.055 mag, and 0.054 mag,
and the reduced $\chi^2$ values are 0.64, 0.67, and 0.62, 
respectively. It is clear from both the RMS and the reduced
$\chi^2$ that the three fits do not differ significantly
from one another. Thus, unless the UVOT photometric precision can be
significantly improved, our data suggest that no color-term
corrections for the UVOT $B$ and $V$ bands are necessary.

The results for the $(U - V)$ color are shown in Figure 15. 
Three fitting function are overplotted: the solid line is
$(U - V)$ = $(u - v)$, which is so far used in this paper; the dashed line is  
$(U - V) = 0.057 + 0.8926 (u - v) + 0.0274 (u - v)^2$, 
which is adopted from Eq. 10 but shifted up and down to
best match the data; and the dash-dotted line is
our best fit to the data, $(U - V) = 0.031 + 0.9150 (u - v) + 0.028 (u - v)^2$. 
The RMS values for the fits are 0.089 mag, 0.081 mag, and 0.079 mag,
and the reduced $\chi^2$ values are 3.39, 1.31, and 1.09, respectively. 
Although adopting a color term for the $(U - V)$ color (the
dashed and dash-dotted lines) decreases the RMS and the $\chi^2$ 
of the fit, we note that the blue color end is dominated by 
just one data point from the Landolt standard star 95\_42, 
and more data for stars with colors $-1.5 < (u - v) < 0.4$ mag
are needed to better constrain the coefficients. When the data for 95\_42
are excluded from the fit, the three fits 
give RMS/(reduced $\chi^2$) values of 0.079 mag/(1.11), 0.082 mag/(1.31),
and 0.080 mag/(1.08), respectively. Both the RMS and the reduced
$\chi^2$ are very similar for the three fits. 

To summarize for the $(U - V)$ color, no color term correction
is necessary when $(u - v) \geq 0.4$ mag. When $(u - v) < 0.4$ mag,
we adopt the best-fit color terms (dash-dotted line in Figure 15).
The corresponding correction in the $U$-band photometry is 
\begin{equation}
\Delta = 0.031 - 0.085 (u - v) + 0.028 (u - v)^2.
\end{equation}

\subsection{Photometry Recipe for the UVOT $UBV$ Filters}

In this section we summarize the key parameters and steps
involved in performing photometry on UVOT $UBV$ filters.
We list all the important parameters required
to do photometry in the IRAF/DAOPHOT package (specifically
for the program ``phot"), but these should be easily exported
to other photometry programs. 

\begin{enumerate}

\item{Input all necessary keywords to the datapars of 
``phot," such as ``EXPOSURE" as the FITS keyword for 
the exposure time.} 

\item{The sky region is defined as an annulus with an inner radius
of 35 pixels (17$\arcsec$.5) and an outer radius of 
45 pixels (22$\arcsec$.5). That is, in ``fitskypars,"
annulus = 35, and dannulus = 10. For 2$\times$2 binned
data, the inner radius becomes 17.5 pixels (17$\arcsec$.5)
and the outer radius 22.5 pixels (22$\arcsec$.5). 
The preferred sky-fitting algorithm is ``mean."}

\item{The photometry aperture is 5 pixels (2$\arcsec$.5)
for unbinned data, and 3 pixels (3$\arcsec$.0) for 
$2\times2$ binned data.}

\item{The photometric zero points are $ZP(U) = 18.24$,
$ZP(B) = 18.92$, and $ZP(V) = 17.69$ mag. The uncertainties for these
zero points are 0.01 mag. The photometry measured from this 
step are denoted by $u$, $b$, and $v$: }

\item{Apply the C-loss correction. The UVOT magnitudes after the 
correction is denoted as $u_c$, $b_c$, and $v_c$. 

$u_c = -2.5 \log_{10} \mu_u + 13.43$, where $\mu_u = -\ln\, ( 1 - 10^{-0.4(u - 13.43)}),$ 

$b_c = -2.5 \log_{10} \mu_b + 14.16$, where $\mu_b = -\ln\, ( 1 - 10^{-0.4(b - 14.16)}),$ 

$v_c = -2.5 \log_{10} \mu_v + 12.92$, where $\mu_v = -\ln\, ( 1 - 10^{-0.4(v - 12.92)}).$ 

}

\item{Apply the color-term correction, and derive the 
standard Johnson/Cousins $UBV$ magnitudes:

$U =\left\{ \begin{array}{ll}
  u_c   & \mbox{ if $(u_c - v_c) \geq 0.4$ mag} \\
  u_c + \Delta   & \mbox{ if $(u_c - v_c) < 0.4$ mag,} 
  \end{array}\right.$

where $\Delta = 0.031 - 0.085 (u_c - v_c) + 0.028 (u_c - v_c)^2,$

$B = b_c, $

$V = v_c. $

}

\end{enumerate}

\section{UVOT Photometry of SN 2005am}

In \S~3, we used the observations of the SN 2005am and SN 2005cf
field stars and the Landolt standard stars, and derived
optimal parameters and the C-loss corrections to
conduct photometry of UVOT images. 
In theory, we can simply follow the recipe in \S~3.6 and 
do photometry of SN 2005am in the UVOT images. However, 
the complex background around the SN (as shown in Figure 1) 
requires special attention when performing photometry.

As discussed in \S~2, in ground-based photometry we used the
PSF-fitting technique to fit for the background around the SN and its
neighboring bright star, and measured their fluxes 
simultaneously. Since the PSF is constant across the image
and there are plenty of bright and isolated stars, a robust
PSF can be constructed, and the overall photometry shows a high
degree of self-consistency; see, for example, the final
light curves shown in Figure 2. In UVOT images, however,
since the intrinsic PSF varies, PSF fitting is not possible without
a precise description of the intrinsic PSFs.  We note that
the UVOT observations of the SN 2005am field could be used to
study the intrinsic PSFs, but to better constrain the PSF
variation dependence on source brightness and position on the
chip, many more observations will be needed than those currently
available to us.  Therefore, we need to resort to aperture photometry
to measure the magnitudes of SN 2005am. 

Because of the complex background around SN 2005am, it is vital
to use a small aperture for photometry, so that the relatively 
poor sky-background estimate has less impact on the final
photometry. Fortunately, our optimal aperture (5 pixels
for unbinned data, and 3 pixels for $2\times2$ binned data)
is small, and is used to derived photometry for SN 2005am.
We also adopt all the other photometric parameters derived
from the previous section.

In \S~3.2.4 we studied several sky-fitting algorithms, and preferred
``mean" for uncrowded SN 2005am field stars. 
However, since our defined sky background region for SN 2005am is
affected by its host-galaxy emission, ``mean" will almost
certainly overestimate the background contribution inside
the aperture radius. We thus performed photometry on SN 2005am using
several sky-fitting algorithms (mean, mode, centroid, 
median, ofilter, and crosscor), averaging the 
final results and also calculating their RMS. 

The final photometry of SN 2005am after the C-loss and color-term 
corrections is listed in Column 3 in Table 11. 
Because we did not include obs4 $U$
(due to its short exposure) and obs3 $V$ (due to its
erroneous exposure time) in the zero-point determination,
we did not perform direct photometry of these images. 
The uncertainties of the photometry are the quadrature sums of 
the zero-point error, the photometric error from
``phot," and the RMS of different sky-fitting methods.

Since we have good $B$ and $V$, and preliminary $U$, calibrations
for part of the SN 2005am UVOT field from the ground, we
can also choose to do differential photometry between the 
SN and the local standard stars, as listed in Column 4 of Table 11.
The UVOT photometric measurements for SN 2005am and the local standard
stars are corrected for the C-loss and color-term corrections
before performing the differential photometry.
The photometric uncertainties are the quadrature sums
of the ``phot" error, the RMS of the different sky-fitting 
methods, and the RMS of the photometry when compared to 
all the local standard stars. The zero point and its uncertainty
are irrelevant in doing differential photometry.
For obs4 $U$ and obs3 $V$, differential photometry is the 
only way to measure the magnitudes for SN 2005am. 

Comparison of the photometry for SN 2005am between the two 
methods reveals no significant difference. We adopt the 
differential photometry as it provides measurements for 
all the images. In Column 5 of Table 8, we list the ground-based estimates
for the magnitudes of SN 2005am at the time of the UVOT observations
by fitting a smooth spline3 (order 2) curve to the the KAIT/Nickel
data after JD 2453450, while Column 6 lists the difference between
our adopted UVOT photometry of SN 2005am (Column 4) and the 
ground-based estimate.  We also combine the
UVOT photometry of SN 2005am with the ground-based Lick light curves
in Figure 16, and redo the MLCS2k2 analysis. Overall, 
the UVOT photometry agrees well with the ground-based estimates,
especially for the observations taken in the first 4 sequences. 
However, the difference in the obs5 $B$ observation, $-0.43\pm0.15$ mag,
is quite large. The SN was not well detected in this
particular image, and the magnitude may be more seriously 
affected by the neighboring bright object than in other frames
as a result of the short exposure time and poor detection. 
The parameters in the MLCS2k2 fit for the combined data set do not
show any significant changes.

\section{UVOT Photometry of GRB 050603}

Since {\it Swift} is designed to study GRBs and related 
phenomena, in this section we analyzed the UVOT observations
of GRB 050603 as a test case for our photometric calibration results.
GRB 050603 was recorded by {\it Swift}/BAT at 6:29:05 on
2005 June 3, and was initially reported as a short GRB (Retter 
et al. 2005), but was revised to be a long GRB lasting about
10~s (Gotz \& Mereghetti 2005). The optical afterglow (OA) was
first identified by Berger \& McWilliam (2005) in images taken
with the du Pont 2.5-m telescope at Las Campanas Observatory at
3.4 hours after the burst. The GRB was subsequently detected at
radio (Cameron 2005) and submillimeter (Barnard et al. 2005)
wavelengths. A tentative redshift of $z = 2.821$ was reported by 
Berger \& Becker (2005). 

{\it Swift}/UVOT began observations of GRB 050603 at 15:42:59
on 2005 Jun.\ 3, $\sim$ 9.2 hours after the burst. Brown et al. 
(2005b) reported a 3.6 mag decline within 2 hours of the UVOT 
observations, though this was debated by Berger (2005). A revised
report for the UVOT observations of GRB 050603 was announced by 
Brown et al. (2005c). 

We retrieved the level-2 UVOT data for GRB 050603 from the {\it Swift}
quicklook archive. The data were observed only in the $V$ band,
and only those observations obtained in the first several days after the burst
were analyzed. These data are organized in two sequences, 
sw00131560001uvv (11 individual exposures) and sw00131560002uvv
(110 individual exposures). Since the OA is faint, we only studied
the images with long exposure times. 

We closely follow the UVOT photometry recipe as described in \S~3.6,
and derive the magnitudes for GRB 050603 as listed in Table 12.
To provide a consistency check, we also measured a relatively bright
and isolated star ``s1" in the GRB 050603 field. A finder chart
for the OA and ``s1" is provided in Figure 17. 

Inspection of the magnitudes of ``s1" (Column 4 of Table 12) reveals
that, although most of the measurements are stable at mag 16.00$\pm$0.04,
some are apparently deviant and show a significant flux deficit. This is
caused by the erroneous exposure times in the image FITS headers as 
discussed in \S~3.2.2.  Following the instructions on the ``UVOT Digest"
page, we listed the corrected exposure times for these images in
Column 5 of Table 12. The magnitudes of the OA (Column 7 of Table 12)
are measured using the corrected exposure times for these images. 

In images when the OA become faint, We coadd several of them to increase the 
S/N. The midpoint for the single and coadded
exposures is calculated as a flux-weighted mean with a power-law
decay index of $-$1.86, as we derive below. 

We show our light curve of GRB 050603 (solid circles) in Figure 18. 
In comparison, we also overplot the reported magnitudes (open circles)
from the {\it Swift} UVOT Team (Brown et al. 2005c), which used the 
preliminary in-flight zero-point calibration as listed in Table 1. 
We fit both datasets with a power-law decay and find the decay
index to be $-1.86\pm$0.06 for the photometry in Table 12 (solid line in
Figure 18), and $-2.01\pm$0.22
for the data from Brown et al. (2005c) (dashed line in Figure 18)\footnote{Brown et al. (2005c)
reported a slightly different power-law decay index ($-1.97\pm$0.22) for
their dataset. The very small difference is probably caused by the 
different methods for calculating the midpoints of the exposures.}. 
The reduced $\chi^2$ for both fits is 2.71 and 2.25, respectively, 
suggesting that either the reported errors for
the magnitudes are underestimated, or the bumps and wiggles around
the fits are real. We find that to achieve a reduced $\chi^2$ of unity,
an additional error of 0.14 mag is required to be added in quadrature
to the photometric error in Table 12. As this is much bigger than the 
photometric scatter after the C-loss correction in the $V$ band
as derived in \S~3.4 (0.045 mag), we suggest that the fluctuations
around the power-law decay fit are real. Similar phenomena have been
observed for other GRBs, such as GRB 021004 (e.g., Lazzati et al. 2002),
GRB 030329 (e.g., Matheson et al. 2003), and perhaps GRB 021211 (Li et al. 2003b).  

Although the two datasets as plotted in Figure 18 are consistent
with each other, and their power-law decay fits have similar indices
and reduced $\chi^2$,  we note that our photometric errors are much
smaller than those reported by Brown et al. (2005c) because of our
tight constraints on the photometry zero points and the C-loss correction.
The uncertainty of our power-law decay index (0.06) is also significantly
smaller than that of the {\it Swift}/UVOT Team (0.22). Thus, our calibration
results provide an improvement to the preliminary in-flight photometric
zero points. 

\section{Discussion}

\subsection{Caveats of Our Photometric Calibration}

Although we have included a limited number of Landolt standard
stars in our analysis, the main source of our photometric
calibration comes from observations of the local
standard stars of two supernovae, SN 2005am and SN 2005cf. Some
unique characteristics of our calibration are as follows:

{(1) The SN field stars extend the calibration to the fainter end. As can be seen 
in Figures 12 and 13, the Landolt stars are mostly
brighter than mag 16, while the SN field stars extend the 
calibration to mag 19. As the majority of GRB optical afterglows are expected to 
be detected much fainter than mag 16, it is critical to study the
photometric calibration at the fainter end. Moreover, the fainter end 
provides the anchor point to determine the zero point in the C-loss
correction as in \S~3.4.}

{ (2) There are many local standard stars in each SN frame.
Consequently, the images can be studied one frame at a time, 
each of which already provides enough information on photometric
consistency between the UVOT photometry and the ground-based 
calibration. The Landolt stars, on the other hand, have a lower
density in the UVOT images, especially since many of them are 
bright and require C-loss correction. It is thus necessary
to combine multiple images to get enough statistics, which could
introduce unexpected variables that hide the true correlations.}

On the other hand, some limitations of our calibration are as follows:

{(1) The SN field stars do not have the photometric precision of the 
Landolt stars. The uncertainties of the SN field stars are mostly
0.02--0.03 mag, while those for the Landolt stars are smaller than 0.01 
mag. In particular, there is only a one-time calibration for the 
$U$-band of the SN 2005cf field, and a preliminary $U$-band calibration
for the SN 2005am field that has not been used in the final calibration.}

{(2) The color range of the dataset is somewhat limited. 
Except for one Landolt star (95\_42) that is quite blue, all
the other stars have $(B - V) > 0.4 $ mag and $(U - V) > 0.4 $ mag. 
More relatively blue stars should be observed and analyzed to better constrain
the color terms of the filters, particularly the $U$ band. 
}

Because of these limitations of our calibration, it is envisioned 
that the calibration results can be further refined
in future studies, although we expect the main improvement to 
be a better constrained color term for the $U$ band. We note that 
we have already derived tight constraints on the photometric zero
points ($\sigma \approx 0.01$ mag) and $m_\infty$ ($\sigma \approx 0.02$ mag),
the saturation magnitude used in the C-loss correction. 

We also note that we used aperture photometry to 
analyze all the images, and it is conceivable that when software
packages have been developed to do intrinsic PSF fitting 
in the UVOT images, the photometric zero points would change,
and better precision could be achieved. 

Our study has not explored all of the possible parameters
for UVOT observations, such as different binning choices,
extremely short and long exposures, etc. Moreover, we have not correlated
the residuals with all possible parameters of UVOT and 
the detectors. 

We emphasize that our photometric zero points and 
optimal photometric parameters may not work well on short ($<$ 20~s) 
UVOT exposures, as suggested by obs4 $U$. Since it is not always
possible to take relatively long exposures with UVOT, it is
important to analyze more UVOT images with short exposures to
establish their photometric calibrations and optimal photometric
parameters.  Unfortunately, it is expected that the photometric
calibration for the short exposures suffers from larger uncertainties
than those derived in this study, as a shorter exposure time 
generally means lower S/N. Observing brighter stars
in the short exposures to increase the S/N is not an 
option due to the effect of coincidence loss. 

We also note that we did not include obs3 $V$ in our 
photometric zero-point determinations due to the erroneous 
exposure time in this image. \emph{Since a small fraction of
UVOT images are affected by the exposure time anomaly, and not all of
them are recognized and documented, 
users are urged to seek consistency checks when
performing photometry on the UVOT images.}
For example, differential photometry should be 
performed whenever possible, either instead of the 
absolute photometry or as a consistency check. When
multiple observations of the same field are available,
photometric consistency should be checked with some bright
field stars, as we did with ``s1" in \S~5 for GRB 050603. 

\subsection{Investigating the Scatter in the UVOT Photometry}

It is a bit disappointing that the photometric
precision can only be achieved to the 0.04 mag level in our 
study of UVOT data (\S~3.4). In comparison, ground-based CCD observations
with a moderate-sized telescope can easily reach a precision of 0.02
mag or better for Landolt standard stars.  The scatter in the
UVOT photometry is unlikely to be caused by the photometric
zero-point errors, as an error in the zero point will cause a
constant offset, not scatter. The lower panels of Figures 11 to
13 suggest that the dispersion in the photometry is present at all
source brightnesses; thus, the scatter is probably not all due to 
photon statistics; rather, part may be intrinsic to the
aperture photometry of the UVOT observations. 

We note that all raw UVOT images contain systematic modulo-8 
fixed-pattern noise (``mod-8 noise'' hereafter) as a result
of pixel subsampling on the 
detector. The CCD detector of UVOT has a physical dimension of 
$385\times288$ pixels, $256\times256$ of which are usable for 
science observations. The detector attains a large format through
a centroiding algorithm to the incoming photons by subsampling 
each physical pixel into $8\times8$ virtual pixels, thus providing
an image of $2048\times2048$ virtual pixels. This subsampling
process introduces faint residuals with a fixed pattern 
(the mod-8 noise) in the raw UVOT images,  which are removed by 
ground processing (rather than by in-orbit {\it Swift} processing). 
The level-2 images we analyzed in our study 
have been processed by the {\it Swift} UVOT pipeline, with the mod-8
noise removed. However, as suggested by the {\it Swift} 
manual, photometric accuracy is destroyed after removing the 
fixed-pattern noise, and flux is conserved only within each 
$8\times8$ pixel block. Moreover, the fixed-pattern noise is 
modified around bright sources, and cannot be recovered without
a well-calibrated Monte Carlo analysis. 

It is possible that the dispersion in our aperture photometry is
caused by the reduced photometric accuracy after removing the 
mod-8 noise. Unfortunately, it is impractical to 
bin the level-2 images by $8\times8$ so that the photometric accuracy 
can be recovered. The pixel scale would be about 4$\arcsec$ 
after binning, and the photometry would be severely affected by 
undersampling the PSF of the stars. 

To further investigate the effect of the mod-8 noise and its
removal in the UVOT photometry, we downloaded the raw images
of obs1 $V$ and experimented with them using the latest HEAsoft 6.0
software supplied by HEASARC. 
We kept the images in the detector pixel frame, 
to avoid possible effects introduced by converting them
to the sky coordinate image using
``uvotxform."\footnote{We have performed the same tests on 
the sky-coordinate images and the results do not change
significantly, suggesting that ``uvotxform" is not the cause 
of the photometric scatter.} ``Uvotbadpix" was performed to
remove bad pixels, ``uvotmodmap" was used to remove the 
mod-8 noises, and ``uvotflatfield" was used to remove pixel-to-pixel
variations in the image due to detector sensitivity. 
We also skipped the ``uvotmodmap" step and generated an image
with the mod-8 noise still included. A difference image
was then generated by subtracting the final image with 
the mod-8 noise removed from the one that skipped ``uvotmodmap,"
and it represents the total effect on the UVOT images
before and after the mod-8 noise is removed by the 
{\it Swift} pipeline. In effect, the difference image is the
mod-8 noise image
normalized by the flat field used in ``uvotflatfield."

We studied the normalized mod-8 noise image that was
applied to obs1 $V$ during our reductions using HEAsoft 6.0.
Visual inspection indicates that the image is very flat.
When viewed with a large contrast, a faint, large-scale
pattern is revealed. When viewed with a small contrast,
residuals that correspond to the stars detected in the 
obs1 $V$ images are apparent. We randomly selected 
10,000 positions in this image, and summed the flux
inside a 5-pixel radius region (the optimal photometric
aperture derived from our study) at each location. 
The histogram of the fluxes from these measurements
shows a Gaussian distribution, with a dispersion of only
0.325 counts ($1\sigma$). We also collected the flux inside a
5-pixel radius region centered on each local standard
star in the SN 2005am field. The total flux for the
SN 2005am  field stars on the normalized mod-8 noise image
has a range from $-7.74$ to $+2.67$ counts. Since the total
flux for the SN 2005am field stars using a 5-pixel photometry
aperture has a range of 175--4400 counts, the effect
of either the random flux fluctuation ($\sigma = 0.325$ counts)
or the local flux fluctuation ($-7.74$ to $+2.67$ counts)
on the final photometry is smaller than 0.005 mag, which 
cannot account for the photometric scatter of 0.04--0.08 mag.

Our analysis also suggests that the mod-8 noise removal procedure
as implemented in HEAsoft 6.0 has negligible effect on the 
final photometry of most stars. Only the faintest stars
that are close to the detection limit will suffer from the 
flux fluctuation of $\sigma = 0.325$ counts (the flux fluctuations
for the SN 2005am field stars are larger, but they are rather bright
stars). In fact, when the raw image of obs1 $V$ is studied,
it yields nearly the same RMS as the processed image with
mod-8 noise removed. The reasons for this are that (1) there 
are not many bad pixels in the UVOT detector, (2) the flat field
in the current CALDB database has a constant 1 at all pixels,
and (3) the mod-8 noise removal does not change the photometry
significantly. 

The flat fields as provided by the latest CALDB release 
(2005 Aug.\ 12) have a constant 1 at all pixels, i.e., no
flat-fielding is really done to remove the pixel-to-pixel
sensitivity variations across the chip.  If the pixel sensitivity
has a variation of 4--5\% across the chip, it will introduce
an intrinsic scatter on the order of 0.04--0.05 mag to the measured
photometry. We consider the dummy flat fields currently being
used in the {\it Swift} UVOT pipeline as the most
likely cause of the photometric scatter, and urge in-orbit
flat fields to be obtained
by observing relatively bright and blank sky regions,  
or by construction super-sky flat fields from all available
UVOT observations. Our early attempt to construct a super-sky 
flat field for the $V$ band already shows some large-scale
structure and variations, suggesting that proper flat-fielding
is crucial in improving the precision of the UVOT photometry. 

We experimented with artificially smoothing the images by
a small amount. When obs1 $V$ is convolved with an elliptical 
Gaussian function with $\sigma$ = 1.0 and 1.5 pixels, the
photometric RMS can be slightly improved by about 0.005 mag. 
As this is not a dramatic improvement, and smoothing (especially
with a high $\sigma$) will change the zero points at specific
photometric aperture radius, we did not re-analyze all
the data after some artificial smoothing. 

We tried to correlate the scatter in the photometry with 
the coordinates of the stars on the images, but found no
apparent trend. 

It is important to reduce the photometric scatter to
improve the photometric accuracy of UVOT images, either through
finding the cause of the scatter from observations
and subsequent pipeline reductions, or by searching for 
more sophisticated photometric methods than the aperture
photometry method we have employed in our study. 
When the intrinsic PSF variation has been established
from more on-flight observations, for example, the PSF-fitting
technique may be applied and may yield better photometric
precision.

\section{Conclusions}

In this paper we present an empirical determination of the 
optimal photometric parameters to analyze UVOT images using
software tools that are familiar to ground-based optical astronomers.
We consider the effect of the coincidence-loss 
correction based on a theoretically motivated model, and provide the photometric zero points and
their uncertainties in the UVOT $U$, $B$, and $V$ filters.
Our calibration results come from the analysis of 
observations of the local standard stars in the SN 2005am and
SN 2005cf fields, and a limited number of Landolt standard stars.
The main conclusions from our analysis are as follows: 

\begin{enumerate}

\item{The optimal aperture radius to do UVOT photometry,
such that the results are most consistent with the ground-based
calibration, is small.  A radius of 5 UVOT pixels should be used for 
unbinned data, and 3 pixels for the $2\times2$ binned data.
This is 2$\arcsec$.5 and 3$\arcsec$.0 in sky coordinates,
respectively.}

\item{The coincidence-loss correction is important even
at relatively faint levels (mag 16 to 19). Based on a theoretically
motivated model, we
consider the coincidence-loss correction with two parameters,
the photometric zero point (ZP) and the saturation magnitude
($m_\infty$), and derive tight constraints on both parameters.
We derive $ZP(U) = 18.24\pm0.01$ mag, $ZP(B) = 18.92\pm0.01$ mag,
$ZP(V) = 17.69\pm0.01$ mag, $m_\infty(U) = 13.43\pm0.02$ mag,
$m_\infty(B) = 14.16\pm0.02$ mag, and $m_\infty(V) = 12.92\pm0.02$ mag.}

\item{With proper coincidence-loss correction, reliable
photometry can be achieved for stars as bright as 
$U$ = 12.4 mag, $B$ = 13.2 mag, and $V$ = 12.0 mag.}

\item{There is a scatter on the order
of 0.04--0.08 mag in the final aperture UVOT photometry
that cannot be easily accounted for, but is likely
to be due to the variation in the pixel sensitivity
for the UVOT detectors. }

\item{The color terms of the UVOT $B$ and $V$ are 
small, and need not to be considered unless the UVOT photometric
precision is significantly improved. The $U$ band needs
to be corrected for color terms when the object has a blue
color [$(U - V) < 0.4$ mag].}

\item{In \S~3.6 of this paper, we offer a step-by-step 
photometry procedure for UVOT images, including all the 
optimal photometric parameters, the photometric zero
points, and the proper coincidence-loss correction.}

\item{We performed photometry of SN 2005am in the UVOT 
images, and compared the results with those from ground-based observations.
The UVOT photometry is generally consistent with the 
ground-based observations, but the difference increased to
$\sim$ 0.5 mag in one measurement when the SN became faint.
Part of the cause for this large difference is the complex
background region around SN 2005am. }

\end{enumerate}

Based on our study of the photometry of UVOT images,
we offer the following suggestions for future {\it Swift}/UVOT
observations and calibrations. We advise that on-board binning 
be avoided for the UVOT observations. Though we were
able to analyze the $2\times$2 binned data,
the binning introduces yet another variable in the 
uncertainties of the photometric zero points.
Many photometric calibration observations 
should be performed, not only of bright Landolt stars,
but also of stars at the fainter end, perhaps to an even
fainter level than we have studied in this paper ($V \approx 18$ mag).
These calibrations should also be done with exposure times
that span a large range, including very short durations ($<$ 20~s).
Observations of standard stars with blue colors should be 
obtained to better constrain the color terms of the filters. 
A series of observations of the same object should be obtained
by varying the pointing so the object is detected at
different positions on the chip, to better constrain the 
uncertainty caused by the PSF variation across the chip. 
We urge in-orbit flat fields/sensitivity maps to be constructed 
and implemented in the data reduction pipeline to increase the
photometric accuracy.

\acknowledgments

The work of A.V.F
is supported by National Science Foundation grant AST-0307894, 
NASA/{\it Swift} grant NNG05GF35G, and the Miller Institute for
Basic Research in Science (U.C. Berkeley; Miller Research Professorship).
SJ gratefully acknowledges support from a Miller Research Fellowship.
KAIT was made possible by generous donations from Sun
Microsystems, Inc., the Hewlett-Packard Company, AutoScope Corporation, Lick
Observatory, the National Science Foundation, the University of California, and
the Sylvia \& Jim Katzman Foundation.

\newpage

\renewcommand{\baselinestretch}{1.0}

\newpage

\begin{deluxetable}{ccccccc}
\tablecaption{Zero points (ZP; mag) for various {\it Swift}/UVOT filters from the {\it Swift} CALDB}
\tablehead{
\colhead{Filter}&\colhead{$U$}&
\colhead{$B$}&\colhead{$V$}&
\colhead{$UVW1$} & \colhead{$UVW2$} &\colhead{$UVM2$} 
}
\startdata
ZP & 18.38 & 19.16 & 17.88 & 17.69 & 17.77 & 17.29 \\
Error & 0.23 & 0.12 & 0.09 & 0.02 & 0.02 & 0.23 \\
Aperture (pixels) & 12 & 12 & 12 &24 & 24 & 24 \\
\enddata    
\end{deluxetable}

\newpage

\begin{deluxetable}{ccccccccccc} 
\tablecaption{Photometry of local standard stars in the field of SN 2005am}
\tablehead{
\colhead{ID}&\colhead{$U$}& \colhead{$N_U$}&
\colhead{$B$}&\colhead{$N_B$} & \colhead{$V$}&\colhead{$N_V$} &
\colhead{$R$} &\colhead{$N_R$}& \colhead{$I$} &\colhead{$N_I$} 
}
\startdata
 1 & $-$ & $-$ & 17.62(01) & 2 & 16.71(03) & 3 & 16.15(01) & 2 & 15.82(02) & 2\\
 2 & 16.80(05) & 1 & 16.90(01) & 2 & 16.39(02) & 3 & 16.08(02) & 2 & 15.75(02) & 2\\
 4 & 16.78(05) & 1 & 15.59(01) & 2 & 14.55(02) & 2 & 13.73(01) & 2 & 13.24(01) & 2\\
 5 & 17.16(05) & 1 & 17.14(01) & 2 & 16.56(02) & 3 & 16.13(02) & 2 & 15.79(02) & 3\\
 6 & 12.39(05) & 1 & 12.30(02) & 4 & 11.99(01) & 3 & 11.78(03) & 3 & 11.58(02) & 2\\
 7 & $-$ & $-$ & 18.20(01) & 2 & 17.22(03) & 4 & 16.55(03) & 3 & 16.07(03) & 2\\
 9 & $-$ & $-$ & 17.79(02) & 3 & 17.23(03) & 2 & 16.96(03) & 3 & 16.63(01) & 3\\
10 & $-$ & $-$ & 17.94(01) & 4 & 17.09(02) & 2 & 16.51(01) & 2 & 15.93(01) & 2\\
12 & 16.15(05) & 1 & 15.57(03) & 4 & 14.76(01) & 3 & 14.30(02) & 2 & 13.88(01) & 2\\
13 & $-$ & $-$ & 19.03(03) & 1 & 17.63(03) & 1 & 16.67(02) & 2 & 15.79(02) & 2\\
14 & $-$ & $-$ & 17.38(03) & 3 & 16.53(01) & 2 & 16.12(01) & 3 & 15.75(02) & 4\\
15 & $-$ & $-$ & 16.73(01) & 2 & 15.95(03) & 1 & 15.56(03) & 2 & 15.13(03) & 1\\
16 & $-$ & $-$ & 17.73(01) & 3 & 17.13(02) & 3 & 16.71(03) & 3 & 16.40(02) & 4\\
17 & $-$ & $-$ & 17.43(01) & 2 & 16.77(03) & 1 & 16.39(01) & 2 & 16.03(03) & 2\\
18 & $-$ & $-$ & 17.62(01) & 4 & 16.91(02) & 4 & 16.50(03) & 4 & 16.16(02) & 4\\
19 & $-$ & $-$ & 18.35(02) & 2 & 17.40(02) & 4 & 16.83(02) & 4 & 16.37(03) & 4\\
20 & $-$ & $-$ & 17.31(01) & 4 & 16.55(03) & 3 & 16.12(03) & 3 & 15.71(02) & 4\\
21 & 16.22(05) & 1 & 15.55(01) & 4 & 14.70(02) & 4 & 14.22(02) & 3 & 13.76(02) & 4\\
22 & 15.02(05) & 1 & 14.98(01) & 4 & 14.47(03) & 3 & 14.18(03) & 4 & 13.90(02) & 4\\
23 & $-$ & $-$ & 18.02(01) & 2 & 17.35(02) & 4 & 16.97(02) & 4 & 16.59(01) & 4\\
24 & $-$ & $-$ & 17.48(01) & 2 & 16.54(02) & 4 & 15.93(01) & 4 & 15.45(02) & 4\\
26 & $-$ & $-$ & 16.49(02) & 4 & 15.38(02) & 4 & 14.68(01) & 4 & 14.12(01) & 4\\
27 & $-$ & $-$ & 17.20(03) & 1 & 16.30(02) & 2 & 15.83(01) & 2 & 15.37(01) & 2\\
28 & $-$ & $-$ & 17.60(02) & 2 & 16.98(01) & 2 & 16.60(04) & 2 & 16.23(04) & 2\\
\enddata
\tablenotetext{}{Note: The RMS of each measurement is indicated in parentheses.}
\end{deluxetable}   

\newpage
\renewcommand{\arraystretch}{0.85}

\begin{deluxetable}{ccccccl}
\tablecaption{Lick Observatory photometry of SN 2005am}
\tablehead{
\colhead{JD $-$} &\colhead{$U$}& \colhead{$B$} & \colhead{ $V$ } &
\colhead{$R$} & \colhead{$I$} &\colhead{Tel.}  \\
\colhead{2450000} & \colhead{(mag)} &
\colhead{(mag)} &
\colhead{(mag)} &
\colhead{(mag)} &
\colhead{(mag)} 
}
\startdata
3435.83 & $-$ & $-$ & $-$ & 13.73(03) & 13.90(03)& KAIT \\
3436.76 & $-$ & 13.90(04) & 13.80(04) & 13.70(04) & 13.93(05)& KAIT \\
3438.74 & $-$ & 13.92(02) & 13.75(03) & 13.65(03) & 13.96(04)& KAIT \\
3439.77 & $-$ & 13.98(02) & 13.78(02) & 13.66(03) & 13.97(03)& KAIT \\
3440.74 & $-$ & 14.04(03) & 13.79(03) & 13.69(03) & 14.01(04)& KAIT \\
3441.75 & $-$ & 14.10(02) & 13.81(03) & 13.73(02) & 14.03(02)& Nickel \\
3442.69 & $-$ & 14.18(03) & 13.84(03) & 13.82(04) & 14.15(05)& KAIT \\
3443.76 & $-$ & 14.27(02) & 13.89(03) & 13.90(02) & 14.18(02)& Nickel \\
3444.73 & $-$ & 14.37(03) & 13.95(03) & 13.99(02) & 14.26(03)& Nickel \\
3444.76 & $-$ & 14.43(03) & 13.98(04) & 14.03(03) & 14.32(05)& KAIT \\
3445.77 & $-$ & 14.53(02) & 14.05(02) & 14.10(05) & 14.39(03)& KAIT \\
3446.75 & $-$ & $-$ & 14.12(04) & $-$ & $-$& KAIT \\
3455.75 & $-$ & 15.90(04) & 14.72(03) & 14.39(04) & 14.21(03)& KAIT \\
3460.73 & $-$ & 16.46(02) & 15.17(02) & 14.76(03) & 14.42(04)& KAIT \\
3462.73 & $-$ & 16.59(02) & 15.35(02) & 14.97(03) & 14.61(03)& KAIT \\
3465.70 & $-$ & 16.78(02) & 15.55(04) & 15.21(03) & 14.88(05)& KAIT \\
3466.69 & $17.02(0.08)$ & 16.86(05) & 15.62(04) & 15.21(04) & 14.96(06)& KAIT \\
3467.73 & $-$ & 16.87(03) & 15.65(03) & 15.30(03) & 15.05(04)& KAIT \\
3470.77 & $-$ & 16.99(02) & 15.77(03) & 15.38(05) & 15.09(04)& Nickel \\
3471.68 & $-$ & 17.01(03) & 15.79(04) & 15.47(02) & 15.15(03)& Nickel \\
3471.68 & $-$ & 17.02(03) & 15.81(05) & 15.49(04) & 15.25(05)& KAIT \\
3472.73 & $-$ & 17.01(03) & 15.91(03) & 15.57(04) & 15.39(04)& KAIT \\
3474.71 & $-$ & 17.10(03) & 15.92(04) & 15.64(03) & 15.44(04)& KAIT \\
3477.67 & $-$ & 17.15(05) & 16.03(06) & 15.75(04) & 15.58(06)& KAIT \\
3486.66 & $-$ & 17.40(04) & 16.31(03) & 16.06(03) & 16.02(03)& KAIT \\
3492.67 & $-$ & $-$ & $-$ & 16.30(03) & $-$& KAIT \\
\enddata    
\tablenotetext{}{Note: Uncertainties are indicated in parentheses;
these are quadrature sums of the PSF-fitting
photometry and the transformation scatter (RMS) from the local standard
stars.}
\end{deluxetable}

\newpage
\renewcommand{\arraystretch}{1.00}

\begin{deluxetable}{lllclr}
\tablecaption{Journal of {\it Swift}/UVOT observations of SN 2005am in $UBV$}
\tablehead{
\colhead{Data ID} & \colhead{Obs ID} &\colhead{Date}
&\colhead{Filter}& \colhead{UT Start} &
\colhead{Exp. time(s)} 
}
\startdata
     &             &           & $U$ &11:38:44 & 201.76 \\
obs1 &sw00030010070&2005-04-04 & $B$ &11:42:14 & 169.41\\
     &             &           & $V$ &11:18:09 & 201.77\\ 
\hline
     &             &           & $U$ & 08:39:02&  209.77\\
obs2\tablenotemark{a} &sw00030010071&2005-04-06 & $B$ &08:42:39 & 144.23\\
     &             &           & $V$ &08:28:12 & 209.77\\
\hline
     &             &           & $U$ &02:23:41 & 82.78\\
obs3 &sw00030010072&2005-04-10 & $B$ &02:25:10 & 40.61\\
     &             &           & $V$ &02:15:04 & 82.77\\
\hline
obs4 &sw00030010073&2005-04-22 & $U$ &02:24:07 & 18.02\\
     &             &           & $V$ &02:08:03 & 157.78\\
\hline
obs5 &sw00030010076&2005-05-17 & $U$ &03:22:59 & 72.78\\
     &             &           & $B$ &03:25:04 & 46.68\\
     &             &           & $V$ &03:15:29 & 72.77\\
\enddata
\tablenotetext{a}{These data are binned 2$\times$2.}
\end{deluxetable}

\newpage

\begin{deluxetable}{ccc|ccc}
\tablecaption{Comparison of different sky background fitting algorithm for obs1 $V$}
\tablehead{
\colhead{} & \multicolumn{2}{c}{Without $CF$}
& \multicolumn{3}{c}{With $CF$} \\
\hline
\colhead{Algorithm} &\colhead{ZP}&\colhead{$\sigma$(ZP)} &
\colhead{$CF$} & \colhead{ZP}&\colhead{$\sigma$(ZP)}
}
\startdata
mean&17.68&0.067&1.046&16.95&0.043\\
mode&17.69&0.074&1.060&16.77&0.044\\
median&17.69&0.072&1.060&16.76&0.043\\
centroid&17.73&0.094&1.090&16.38&0.049\\
ofilter&17.69&0.073&1.060&16.76&0.043\\
crosscor&17.70&0.078&1.070&16.63&0.044\\
\enddata
\end{deluxetable} 

\newpage


\begin{deluxetable}{ccccccc} 
\tablecaption{Photometry of local standard stars in the field of SN 2005cf}
\tablehead{
\colhead{ID}&\colhead{$U$}& \colhead{$N_U$}&
\colhead{$B$}&\colhead{$N_B$} & \colhead{$V$}&\colhead{$N_V$} 
}
\startdata
1 & $-$ & $-$ & 15.27(01) & 3 & 14.38(01) & 3 \\
2 & 13.65(03) & 1 & 13.49(01) & 3 & 12.80(01) & 3 \\
3 & 16.37(03) & 1 & 15.62(01) & 4 & 14.68(01) & 3 \\
4 & 16.58(03) & 1 & 16.47(01) & 3 & 15.76(01) & 3 \\
5 & 15.31(03) & 1 & 15.33(01) & 4 & 14.82(01) & 4 \\
6 & 14.01(03) & 1 & 14.06(02) & 5 & 13.60(01) & 4 \\
7 & 18.30(03) & 1 & 18.43(01) & 4 & 17.79(01) & 2 \\
8 & 18.93(03) & 1 & 17.81(02) & 5 & 16.26(01) & 4 \\
9 & 15.83(03) & 1 & 15.66(02) & 5 & 14.99(01) & 4 \\
10 & 18.29(03) & 1 & 17.12(01) & 5 & 15.95(01) & 3 \\
11 & 14.96(03) & 1 & 14.75(01) & 5 & 14.02(01) & 4 \\
12 & 19.35(03) & 1 & 18.34(01) & 4 & 17.33(01) & 4 \\
13 & 15.34(03) & 1 & 14.76(01) & 5 & 13.88(01) & 4 \\
14 & 18.56(03) & 1 & 18.14(02) & 5 & 17.39(02) & 4 \\
15 & 18.30(03) & 1 & 17.59(01) & 5 & 16.72(01) & 4 \\
16 & 17.93(03) & 1 & 17.98(02) & 4 & 17.45(01) & 4 \\
17 & 19.02(03) & 1 & 18.21(03) & 1 & 17.23(03) & 1 \\
18 & 15.37(03) & 1 & 14.97(03) & 1 & 14.16(03) & 1 \\
20 & 16.43(03) & 1 & 15.95(03) & 1 & 15.13(03) & 1 \\
21 & $-$ & $-$ & 18.50(03) & 1 & 17.01(03) & 1 \\
22 & 17.95(03) & 1 & 17.20(03) & 1 & 16.25(03) & 1 \\
23 & 17.36(03) & 1 & 17.32(03) & 1 & 16.70(03) & 1 \\
24 & $-$ & $-$ & 19.06(03) & 1 & 17.55(03) & 1 \\
25 & 14.47(03) & 1 & 14.41(03) & 1 & 13.78(03) & 1 \\
26 & 16.38(03) & 1 & 16.07(03) & 1 & 15.30(03) & 1 \\
27 & 15.40(03) & 1 & 14.56(03) & 1 & 13.55(03) & 1 \\
28 & 17.49(03) & 1 & 17.46(03) & 1 & 16.85(03) & 1 \\
29 & 18.59(03) & 1 & 18.29(03) & 1 & 17.52(03) & 1 \\
30 & 18.76(03) & 1 & 18.49(03) & 1 & 17.75(03) & 1 \\
31 & 18.60(03) & 1 & 18.23(03) & 1 & 17.50(03) & 1 \\
32 & 16.26(03) & 1 & 15.88(03) & 1 & 15.02(03) & 1 \\
33 & 15.82(03) & 1 & 15.46(03) & 1 & 14.68(03) & 1 \\
34 & $-$ & $-$ & 18.71(03) & 1 & 17.13(03) & 1 \\
35 & 18.44(03) & 1 & 18.39(03) & 1 & 17.78(03) & 1 \\
36 & 14.56(03) & 1 & 14.51(03) & 1 & 13.93(03) & 1 \\
37 & 17.93(03) & 1 & 18.04(03) & 1 & 17.46(03) & 1 \\
39 & 16.65(03) & 1 & 16.27(03) & 1 & 15.53(03) & 1 \\
40 & $-$ & $-$ & 18.53(03) & 1 & 17.57(03) & 1 \\
\enddata
\tablenotetext{}{Note: Only $U$, $B$, and $V$ magnitudes are 
reported. The RMS of each measurement is indicated in parentheses.}
\end{deluxetable}   

\newpage
\small

\begin{deluxetable}{llllr}
\tablecaption{{\it Swift}/UVOT observations of SN 2005cf analyzed in this paper}
\tablehead{
\colhead{Obs ID} &\colhead{Date}
&\colhead{Filter}& \colhead{N(image)\tablenotemark{a}} &
\colhead{Exp. time(s)\tablenotemark{b}} 
}
\startdata
sw00030028007 &  2005-06-04 & U & 1 & 71.77 \\
sw00030028007 &  2005-06-04 & B & 1 & 71.78 \\
sw00030028007 &  2005-06-04 & V & 1 & 71.78 \\
sw00030028010 &  2005-06-05 & U & 1 & 77.78 \\
sw00030028010 &  2005-06-05 & B & 1 & 59.24 \\
sw00030028010 &  2005-06-05 & V & 1 & 77.77 \\
sw00030028013 &  2005-06-06 & U & 1 & 77.78 \\
sw00030028013 &  2005-06-06 & B & 1 & 57.65 \\
sw00030028013 &  2005-06-06 & V & 1 & 77.77 \\
sw00030028022 &  2005-06-09 & U & 1 & 91.78 \\
sw00030028022 &  2005-06-09 & B & 1 & 69.78 \\
sw00030028022 &  2005-06-09 & V & 1 & 91.76 \\
sw00030028025 &  2005-06-10 & U & 1 & 85.75 \\
sw00030028025 &  2005-06-10 & B & 1 & 67.78 \\
sw00030028025 &  2005-06-10 & V & 1 & 85.77 \\
sw00030028058 &  2005-06-29 & U & 1 & 71.78 \\
sw00030028058 &  2005-06-29 & B & 1 & 51.63 \\
sw00030028058 &  2005-06-29 & V & 1 & 71.77 \\
sw00030028064 &  2005-07-12 & U & 1 & 70.78 \\
sw00030028064 &  2005-07-12 & B & 1 & 49.16 \\
sw00030028064 &  2005-07-12 & V & 1 & 70.78 \\
sw00030028066 &  2005-07-23 & U & 1 & 439.78 \\
sw00030028066 &  2005-07-23 & B & 1 & 414.01 \\
sw00030028066 &  2005-07-23 & V & 1 & 439.77 \\
\enddata
\tablenotetext{a}{Number of images in the sequence.}
\tablenotetext{b}{Total exposure time in the whole sequence.}
\end{deluxetable}

\newpage

\begin{deluxetable}{llccrr}
\tablecaption{{\it Swift}/UVOT observations of the Landolt fields analyzed in this paper}
\tablehead{
\colhead{Obs ID} &\colhead{Object} &\colhead{Date}
&\colhead{Filter}& \colhead{N(image)\tablenotemark{a}} &
\colhead{Exp. time(s)\tablenotemark{b}} 
}
\startdata
sw00055450008 & SA104NE& 2005-04-05 & V & 1 & 737.52 \\
sw00055400016 & SA104N & 2005-04-19 & V & 1 & 1146.12 \\
sw00055450010 & SA104NE& 2005-04-19 & V & 1 & 697.78 \\
sw00055350013 & SA104SW& 2005-05-10 & V & 1 & 180.04\\
sw00054350014 & SA95SW & 2005-07-07 & U & 5 & 1435.88 \\
sw00054350014 & SA95SW & 2005-07-07 & B & 5 & 1295.40 \\
sw00054350014 & SA95SW & 2005-07-07 & V & 5 & 1435.81 \\
sw00054350015 & SA95SW & 2005-07-08 & U & 13 & 6474.45 \\
sw00054350015 & SA95SW & 2005-07-08 & B & 13 & 6080.93 \\
sw00054350015 & SA95SW & 2005-07-08 & V & 13 & 6473.04 \\
sw00054350016 & SA95SW & 2005-07-11 & U & 29 & 11260.67 \\
sw00054350016 & SA95SW & 2005-07-11 & B & 29 & 10009.79 \\
sw00054350016 & SA95SW & 2005-07-11 & V & 29 & 11276.28 \\
sw00054350017 & SA95SW & 2005-07-09 & U & 3 & 1601.43 \\
sw00054350017 & SA95SW & 2005-07-09 & B & 3 & 1511.64 \\
sw00054350017 & SA95SW & 2005-07-09 & V & 3 & 1601.35 \\
sw00055763001 & SA95-42 & 2005-07-07 & B & 1 & 568.48 \\
sw00055763002 & SA95-42 & 2005-07-07 & V & 1 & 509.65 \\
sw00055763003 & SA95-42 & 2005-07-07 & B & 1 & 569.41 \\
sw00055763004 & SA95-42 & 2005-07-07 & V & 1 & 509.00 \\
\enddata
\tablenotetext{a}{Number of images in the sequence.}
\tablenotetext{b}{Total exposure time in the whole sequence.}
\end{deluxetable}

\newpage


\begin{deluxetable}{lccccrcrcr}
\tablecaption{The average {\it Swift}/UVOT photometry for the 
standard stars\tablenotemark{a,b}}
\tablehead{
\colhead{Name} &\colhead{U} &\colhead{B}
&\colhead{V}& \colhead{u} &\colhead{N(u)}
& \colhead{b} &\colhead{N(b)}
& \colhead{v} &\colhead{N(v)}
}
\startdata
95\_101 & 13.718(011) & 13.455(004) & 12.677(003) & 14.127(017) & 49 & 14.347(023) & 49 & 13.296(014) & 49 \\
95\_15 & 12.171(004) & 12.014(001) & 11.302(001) & 13.472(011) & 2 & 14.123(029) & 2 & 12.874(007) & 2 \\
95\_16 & 16.941(037) & 15.619(020) & 14.313(012) & 16.943(065) & 2 & 15.805(014) & 2 & 14.528(016) & 2 \\
95\_42 & 14.280(011) & 15.391(009) & 15.606(006) & 14.353(013) & 34 & 15.648(026) & 34 & 15.703(025) & 34 \\
95\_43 & 11.297(004) & 11.313(003) & 10.803(002) & 13.467(061) & 29 & 14.275(171) & 29 & 12.881(033) & 29 \\
95\_96 & 10.229(004) & 10.157(002) & 10.010(002) & 13.863(665) & 21 & 15.115(1265) & 21 & 13.166(494) & 21 \\
95\_97 & 16.104(031) & 15.724(023) & 14.818(001) & 16.173(035) & 48 & 15.889(039) & 48 & 14.946(024) & 48 \\
95\_98 & 16.721(018) & 15.629(002) & 14.448(001) & 16.779(033) & 49 & 15.794(032) & 49 & 14.621(021) & 49 \\
104\_335 & 12.432(010) & 12.287(010) & 11.665(010) & $-$ & $-$ & $-$ & $-$ & 12.948(030) & 2 \\
104\_367 & 16.357(037) & 16.483(033) & 15.844(025) & $-$ & $-$ & $-$ & $-$ & 15.900(052) & 2 \\
104\_484 & 16.162(024) & 15.430(020) & 14.406(007) & $-$ & $-$ & $-$ & $-$ & 14.640(018) & 2 \\
104\_485 & 16.348(042) & 15.855(036) & 15.017(011) & $-$ & $-$ & $-$ & $-$ & 15.141(032) & 2 \\
sn05cf\_1 & $-$ & 15.265(010) & 14.380(007) & $-$ & $-$ & 15.470(028) & 8 & 14.558(030) & 8 \\
sn05cf\_2 & 13.650(030) & 13.486(012) & 12.799(013) & 14.078(015) & 8 & 14.332(008) & 8 & 13.341(011) & 8 \\
sn05cf\_3 & 16.370(030) & 15.625(008) & 14.676(012) & 16.428(031) & 8 & 15.751(047) & 8 & 14.766(033) & 8 \\
sn05cf\_4 & 16.582(030) & 16.466(008) & 15.756(007) & 16.577(062) & 8 & 16.501(038) & 8 & 15.805(043) & 8 \\
sn05cf\_5 & 15.311(030) & 15.328(011) & 14.820(011) & 15.437(035) & 8 & 15.491(023) & 8 & 14.901(030) & 8 \\
sn05cf\_6 & 14.014(030) & 14.059(027) & 13.604(014) & 14.338(013) & 8 & 14.585(016) & 8 & 13.866(021) & 8 \\
sn05cf\_7 & 18.297(030) & 18.434(017) & 17.786(001) & 18.215(078) & 8 & 18.406(099) & 8 & 17.783(166) & 8 \\
sn05cf\_8 & 18.926(030) & 17.810(020) & 16.264(007) & 18.925(187) & 8 & 17.788(100) & 8 & 16.213(055) & 8 \\
sn05cf\_9 & 15.831(030) & 15.660(023) & 14.986(010) & 15.858(022) & 8 & 15.799(037) & 8 & 15.035(043) & 8 \\
sn05cf\_10 & 18.294(030) & 17.124(016) & 15.947(011) & 18.277(108) & 8 & 17.142(057) & 8 & 15.951(073) & 8 \\
sn05cf\_11 & 14.956(030) & 14.747(008) & 14.022(005) & 15.081(031) & 8 & 15.044(022) & 8 & 14.219(032) & 8 \\
sn05cf\_12 & 19.353(030) & 18.338(012) & 17.327(017) & 19.305(227) & 8 & 18.342(143) & 8 & 17.300(094) & 8 \\
sn05cf\_13 & 15.340(030) & 14.760(006) & 13.883(006) & 15.424(030) & 8 & 15.056(020) & 8 & 14.088(021) & 8 \\
sn05cf\_14 & 18.558(030) & 18.139(025) & 17.393(027) & 18.298(142) & 8 & 18.093(093) & 8 & 17.375(086) & 8 \\
sn05cf\_15 & 18.297(030) & 17.591(011) & 16.715(007) & 18.172(069) & 8 & 17.611(067) & 8 & 16.708(055) & 8 \\
sn05cf\_16 & 17.933(030) & 17.981(020) & 17.450(015) & 17.902(095) & 8 & 17.992(090) & 8 & 17.409(182) & 8 \\
sn05cf\_17 & 19.016(030) & 18.214(030) & 17.233(030) & 19.079(124) & 8 & 18.330(133) & 8 & 17.301(100) & 8 \\
sn05cf\_18 & 15.370(030) & 14.970(030) & 14.161(030) & 15.479(034) & 8 & 15.203(032) & 8 & 14.362(032) & 8 \\
sn05cf\_20 & 16.435(030) & 15.948(030) & 15.133(030) & 16.443(045) & 8 & 16.024(038) & 8 & 15.224(057) & 8 \\
sn05cf\_21 & $-$ & 18.497(030) & 17.008(030) & $-$ & $-$ & 18.447(186) & 8 & 16.976(091) & 8 \\
sn05cf\_22 & 17.954(030) & 17.200(030) & 16.255(030) & 17.840(081) & 8 & 17.204(084) & 8 & 16.252(077) & 8 \\
sn05cf\_23 & 17.361(030) & 17.322(030) & 16.703(030) & 17.384(082) & 8 & 17.289(067) & 8 & 16.718(068) & 8 \\
sn05cf\_24 & $-$ & 19.058(030) & 17.549(030) & $-$ & $-$ & 19.117(253) & 8 & 17.644(074) & 8 \\
sn05cf\_25 & 14.470(030) & 14.414(030) & 13.780(030) & 14.659(023) & 8 & 14.797(010) & 8 & 14.010(015) & 8 \\
sn05cf\_26 & 16.381(030) & 16.073(030) & 15.297(030) & 16.346(055) & 8 & 16.152(016) & 8 & 15.346(035) & 8 \\
sn05cf\_27 & 15.401(030) & 14.565(030) & 13.546(030) & 15.535(039) & 8 & 14.911(019) & 8 & 13.846(009) & 8 \\
sn05cf\_28 & 17.487(030) & 17.456(030) & 16.853(030) & 17.437(051) & 8 & 17.471(043) & 8 & 16.847(085) & 8 \\
sn05cf\_29 & 18.587(030) & 18.286(030) & 17.516(030) & 18.824(150) & 8 & 18.393(152) & 8 & 17.493(122) & 8 \\
sn05cf\_30 & 18.757(030) & 18.488(030) & 17.754(030) & 18.773(180) & 8 & 18.466(223) & 8 & 17.733(208) & 8 \\
sn05cf\_31 & 18.601(030) & 18.225(030) & 17.504(030) & 18.535(106) & 8 & 18.260(101) & 8 & 17.426(127) & 8 \\
sn05cf\_32 & 16.265(030) & 15.879(030) & 15.016(030) & 16.407(053) & 8 & 15.971(024) & 8 & 15.095(045) & 8 \\
sn05cf\_33 & 15.822(030) & 15.462(030) & 14.682(030) & 15.857(030) & 8 & 15.612(034) & 8 & 14.779(029) & 8 \\
sn05cf\_34 & $-$ & 18.713(030) & 17.127(030) & $-$ & $-$ & 18.694(102) & 8 & 17.191(082) & 8 \\
sn05cf\_35 & 18.440(030) & 18.386(030) & 17.781(030) & 18.598(153) & 3 & 18.433(187) & 3 & 17.862(150) & 3 \\
sn05cf\_36 & 14.560(030) & 14.509(030) & 13.926(030) & 14.757(025) & 3 & 14.901(004) & 3 & 14.148(008) & 3 \\
sn05cf\_37 & 17.929(030) & 18.043(030) & 17.457(030) & 17.930(013) & 2 & 18.052(035) & 2 & 17.401(024) & 2 \\
sn05cf\_39 & 16.651(030) & 16.266(030) & 15.534(030) & 16.633(030) & 2 & 16.367(032) & 2 & 15.573(056) & 2 \\
sn05cf\_40 & $-$ & 18.532(030) & 17.565(030) & $-$ & $-$ & 18.454(076) & 3 & 17.548(039) & 3 \\
sn05am\_1 & $-$ & 17.617(013) & 16.712(027) & $-$ & $-$ & 17.580(064) & 2 & 16.617(104) & 3 \\
sn05am\_2 & $-$ & 16.899(001) & 16.389(019) & $-$ & $-$ & 16.952(035) & 2 & 16.349(049) & 3 \\
sn05am\_4 & $-$ & 15.589(001) & 14.550(016) & $-$ & $-$ & 15.763(025) & 2 & 14.655(022) & 3 \\
sn05am\_5 & $-$ & 17.142(008) & 16.557(017) & $-$ & $-$ & 17.193(013) & 2 & 16.528(044) & 3 \\
sn05am\_6 & $-$ & 12.304(019) & 11.986(012) & $-$ & $-$ & 14.125(045) & 2 & 13.006(015) & 3 \\
sn05am\_7 & $-$ & 18.201(005) & 17.220(028) & $-$ & $-$ & 18.094(037) & 2 & 17.224(047) & 3 \\
sn05am\_9 & $-$ & 17.791(024) & 17.233(025) & $-$ & $-$ & 17.740(112) & 2 & 17.251(129) & 3 \\
sn05am\_10 & $-$ & 17.935(012) & 17.086(017) & $-$ & $-$ & 17.837(117) & 2 & 17.085(038) & 3 \\
sn05am\_12 & $-$ & 15.566(025) & 14.764(012) & $-$ & $-$ & 15.710(008) & 2 & 14.808(030) & 3 \\
sn05am\_13 & $-$ & 19.025(030) & 17.634(030) & $-$ & $-$ & 18.948(018) & 2 & 17.549(032) & 3 \\
sn05am\_14 & $-$ & 17.379(025) & 16.527(005) & $-$ & $-$ & 17.340(051) & 2 & 16.602(104) & 3 \\
sn05am\_15 & $-$ & 16.734(010) & 15.953(030) & $-$ & $-$ & 16.815(035) & 2 & 16.041(039) & 3 \\
sn05am\_16 & $-$ & 17.728(014) & 17.134(019) & $-$ & $-$ & 17.710(086) & 2 & 17.123(086) & 3 \\
sn05am\_17 & $-$ & 17.432(014) & 16.769(030) & $-$ & $-$ & 17.443(025) & 2 & 16.764(026) & 3 \\
sn05am\_18 & $-$ & 17.620(009) & 16.913(024) & $-$ & $-$ & 17.649(087) & 2 & 16.941(062) & 3 \\
sn05am\_19 & $-$ & 18.353(018) & 17.396(018) & $-$ & $-$ & 18.406(173) & 2 & 17.295(005) & 3 \\
sn05am\_20 & $-$ & 17.309(005) & 16.547(027) & $-$ & $-$ & 17.318(035) & 2 & 16.532(033) & 3 \\
sn05am\_21 & $-$ & 15.554(005) & 14.695(020) & $-$ & $-$ & 15.691(020) & 2 & 14.791(052) & 3 \\
sn05am\_22 & $-$ & 14.978(006) & 14.465(026) & $-$ & $-$ & 15.212(005) & 2 & 14.586(026) & 3 \\
sn05am\_23 & $-$ & 18.016(012) & 17.348(021) & $-$ & $-$ & 17.975(093) & 2 & 17.365(099) & 3 \\
sn05am\_24 & $-$ & 17.476(013) & 16.544(017) & $-$ & $-$ & 17.539(083) & 2 & 16.512(071) & 3 \\
sn05am\_26 & $-$ & 16.485(023) & 15.376(023) & $-$ & $-$ & 16.615(086) & 2 & 15.488(077) & 3 \\
sn05am\_27 & $-$ & 17.201(030) & 16.295(021) & $-$ & $-$ & 17.219(019) & 2 & 16.283(053) & 3 \\
sn05am\_28 & $-$ & 17.603(015) & 16.980(009) & $-$ & $-$ & 17.598(063) & 2 & 17.024(092) & 3 \\
\enddata
\tablenotetext{a}{The zero points for the UVOT photometry are $ZP(U)$ = 18.24 mag, $ZP(B)$ = 18.92 
mag, and $ZP(V)$ = 17.69 mag.}
\tablenotetext{b}{The RMS of each measurement is indicated in parentheses.}
\end{deluxetable}

\small

\newpage
\begin{deluxetable}{ll}
\tablecaption{Aperture photometry parameters for {\it Swift}/UVOT\tablenotemark{a}}
\tablehead{
\colhead{Parameter} &\colhead{Value}
}
\startdata
Sky region & 35$-$45 pixels (17.5$-$22.5 for 2$\times$2 binned data) \\
Sky fitting algorithm &  mean (but see text for crowded regions) \\
Aperture size & 5 pixels (3 pixels for 2$\times$2 binned data) \\
ZP($U$)  & 18.24$\pm$0.01 \\
ZP($B$) & 18.92$\pm$0.01 \\
ZP($V$) & 17.69$\pm$0.01 \\
m$_\infty$($U$) & 13.43$\pm$0.02 \\
m$_\infty$($B$) & 14.16$\pm$0.02 \\
m$_\infty$($V$) & 12.92$\pm$0.02 \\
\enddata
\tablenotetext{a}{See \S~3.4 for a complete recipe on how to 
use these parameters to do photometry on UVOT images.}
\end{deluxetable}

\small

\newpage

\begin{deluxetable}{lccccc}
\tablecaption{{\it Swift}/UVOT photometry of SN 2005am}
\tablehead{
\colhead{JD $-$} &\colhead{Filter}& \colhead{Mag1\tablenotemark{a}} & \colhead{Mag2\tablenotemark{b} } &
\colhead{Mag(Lick)\tablenotemark{c}} & \colhead{Diff.\tablenotemark{d}} \\
\colhead{2450000} &  &
\colhead{(mag)} &
\colhead{(mag)} &
\colhead{(mag)} &
\colhead{(mag)} 
}
\startdata
3464.986 & $U$ & 17.11(06) & 17.13(08) &   $-$     &   $-$ \\
3464.989 & $B$ & 16.85(05) & 16.86(06) & 16.73(03) & 0.13(07) \\
3464.973 & $V$ & 15.49(04) & 15.52(05) & 15.50(02) & 0.02(06) \\
3466.862 & $U$ & 17.11(09) & 17.14(11) & 17.02(08) & 0.12(12) \\
3466.864 & $B$ & 16.85(06) & 16.88(09) & 16.86(05) & 0.02(10) \\
3466.855 & $V$ & 15.52(04) & 15.53(05) & 15.62(04) & $-$0.09(06) \\
3470.600 & $U$ & 17.23(09) & 17.28(12) &   $-$     &   $-$ \\
3470.601 & $B$ & 16.97(08) & 16.99(10) & 16.99(02) & $-$0.00(11) \\
3470.594 & $V$ &   $-$     & 15.76(12) & 15.77(03) & $-$0.01(13) \\
3482.600 & $U$ &   $-$     & 17.54(15) &   $-$     &   $-$ \\
3482.590 & $V$ & 16.20(06) & 16.19(09) & 16.18(04) & 0.01(09) \\
3507.641 & $U$ & 18.17(14) & 18.23(15) &   $-$     &   $-$ \\
3507.642 & $B$ & 17.49(08) & 17.51(14) & 17.94(08) & $-$0.43(15) \\
3507.636 & $V$ & 16.84(11) & 16.86(14) & 17.02(06) & $-$0.16(15)  \\
\enddata
\tablenotetext{a}{Magnitude measured following the UVOT 
photometry recipe as described in \S~3.4. }
\tablenotetext{b}{Magnitude measured from differential photometry.}
\tablenotetext{c}{Magnitude measured (or extrapolated) from Lick Observatory observations.}
\tablenotetext{d}{The difference between Column 4 and Column 5.}
\end{deluxetable}


\newpage

\begin{deluxetable}{rrrrrrrr}
\tablecaption{{\it Swift}/UVOT photometry of GRB 050603}
\tablehead{
\colhead{Date} & \colhead{UT}
& \colhead{EXP(s)\tablenotemark{a}} &\colhead{Mag(S1)\tablenotemark{b}} 
&\colhead{EXP(s)$\prime$\tablenotemark{c}}&\colhead{t(mid)(h)\tablenotemark{d}}
&\colhead{Mag(OA)} &\colhead{$\sigma$(Mag)}
}
\startdata
Jun-03& 15:42:59&  209.77& 15.98 & ...   &  9.26& 18.30& 0.13 \\
Jun-03& 15:46:32& 1297.63& 15.97 & ...   &  9.47& 18.19& 0.06 \\
Jun-03& 17:19:24&  209.78& 15.98 & ...   & 10.87& 18.49& 0.14 \\
Jun-03& 17:22:57& 1500.44& 18.89\tablenotemark{e}& 109.83& 10.91& 18.41& 0.17 \\
Jun-03& 18:55:52&  209.76& 15.99 & ...   & 12.48& 18.74& 0.16 \\
Jun-03& 18:59:21& 1772.35& 16.00 & ...   & 12.75& 19.19& 0.08 \\
Jun-03& 20:32:17&  209.77& 15.98 & ...   & 14.08& 19.01& 0.19 \\
Jun-03& 20:35:50& 2103.78& 15.99 & ...   & 14.40& 19.30& 0.09 \\
Jun-03& 22:12:17& 1199.90& 15.99 & ...   & 15.89& 19.37& 0.12 \\
Jun-04& 00:04:22& 1204.78& 15.99 & ...   & 17.76& 19.47& 0.12 \\
Jun-04& 01:25:45& 2055.79& 16.00 & ...   & 19.23& 20.07& 0.14 \\
Jun-04& 03:03:35& 1947.77& 15.99 & ...   & 20.85& 20.13& 0.15 \\
Jun-04& 04:38:45& 2055.78& 16.06 & ...   & 22.45& 20.48& 0.19 \\
\hline
Jun-04& 11:05:35& 1947.78& 15.99 & ...   &      &      &      \\
Jun-04& 12:39:44&  848.78& 15.97 & ...   &      &      &      \\
Jun-04& 14:16:44&  848.78& 16.00 & ...   &      &      &      \\
Combined      &         & 3645.34&  ...  & ...   & 29.84& 20.48& 0.14 \\
\hline
Jun-04& 15:53:04& 1047.79& 16.00 & ...   &      &      &      \\
Jun-04& 17:29:28& 1304.79& 16.04 & ...   &      &      &      \\
Jun-04& 19:05:33& 1355.78& 16.04 & ...   &      &      &     \\
Jun-04& 20:44:10& 1697.79& 17.99\tablenotemark{e}& 277.25&      &      &    \\
Jun-04& 22:19:10& 1097.78& 15.98 & ...   &      &      &    \\
Jun-04& 23:56:40& 1998.78& 17.22\tablenotemark{e}& 614.02&      &      &      \\
Jun-05& 01:32:38& 2004.78& 15.99 & ...   &      &      &      \\
Combined      &         &10507.49&  ...  &7702.17& 38.22& 20.91& 0.11 \\
\hline
Jun-05& 09:35:40& 1997.79& 15.99 & ...   &      &      &      \\
Jun-05& 11:10:40& 1398.78& 15.98 & ...   &      &      &      \\
Jun-05& 12:46:30&  697.78& 16.01 & ...   &      &      &      \\
Jun-05& 14:22:44&  848.77& 15.97 & ...   &      &      &      \\
Jun-05& 16:00:05& 1047.77& 16.01 & ...   &      &      &      \\
Jun-05& 17:35:20& 1255.68& 16.09 & ...   &      &      &      \\
Jun-05& 19:12:53& 1555.77& 19.02\tablenotemark{e}& 119.02&      &      &      \\
Combined      &         & 8802.34&  ...  &7365.59& 54.63& 21.75& 0.23 \\
\hline
Jun-05& 20:50:15& 1747.78& 16.00 & ...   &      &      &      \\
Jun-05& 22:27:34& 1948.78& 15.99 & ...   &      &      &      \\
Jun-06& 00:03:40& 1998.78& 16.35\tablenotemark{e}&1319.79&      &      &      \\
Jun-06& 01:38:43& 2055.77& 15.98 & ...   &      &      &      \\
Jun-06& 03:15:18& 2025.79& 16.05 & ...   &      &      &      \\
Combined      &         & 9776.90&  ...  &9097.91& 65.82& $>$22.21& 0.32 \\
\enddata
\tablenotetext{a}{Original exposure time (in seconds) in the FITS file.}
\tablenotetext{b}{Magnitude of S1. Most of the measurements are stable 
at mag 16.00$\pm$0.04, but some are apparently deviant.}
\tablenotetext{c}{Corrected exposure time; see text for details. }
\tablenotetext{d}{Flux-weighted midpoint of the exposure in hours.}
\tablenotetext{e}{Apparent flux deficit for S1.}
\end{deluxetable}

\newpage

\small
\clearpage

\begin{figure}
\caption{Finder charts for the field of SN 2005am. The left panel shows
a KAIT $R$-band image ($6\arcmin.6\times6\arcmin.6$) taken on 2005
Mar.\ 13, and the right panel shows a Nickel $R$-band image
($6\arcmin.3\times6\arcmin.3$) taken on 2005 Mar.\ 12. North is up and east
is to the left. SN 2005am and the local standard stars listed in Table 2
are labeled.}
\label{1}
\end{figure}

\newpage
\clearpage

\begin{figure}
{\plotfiddle{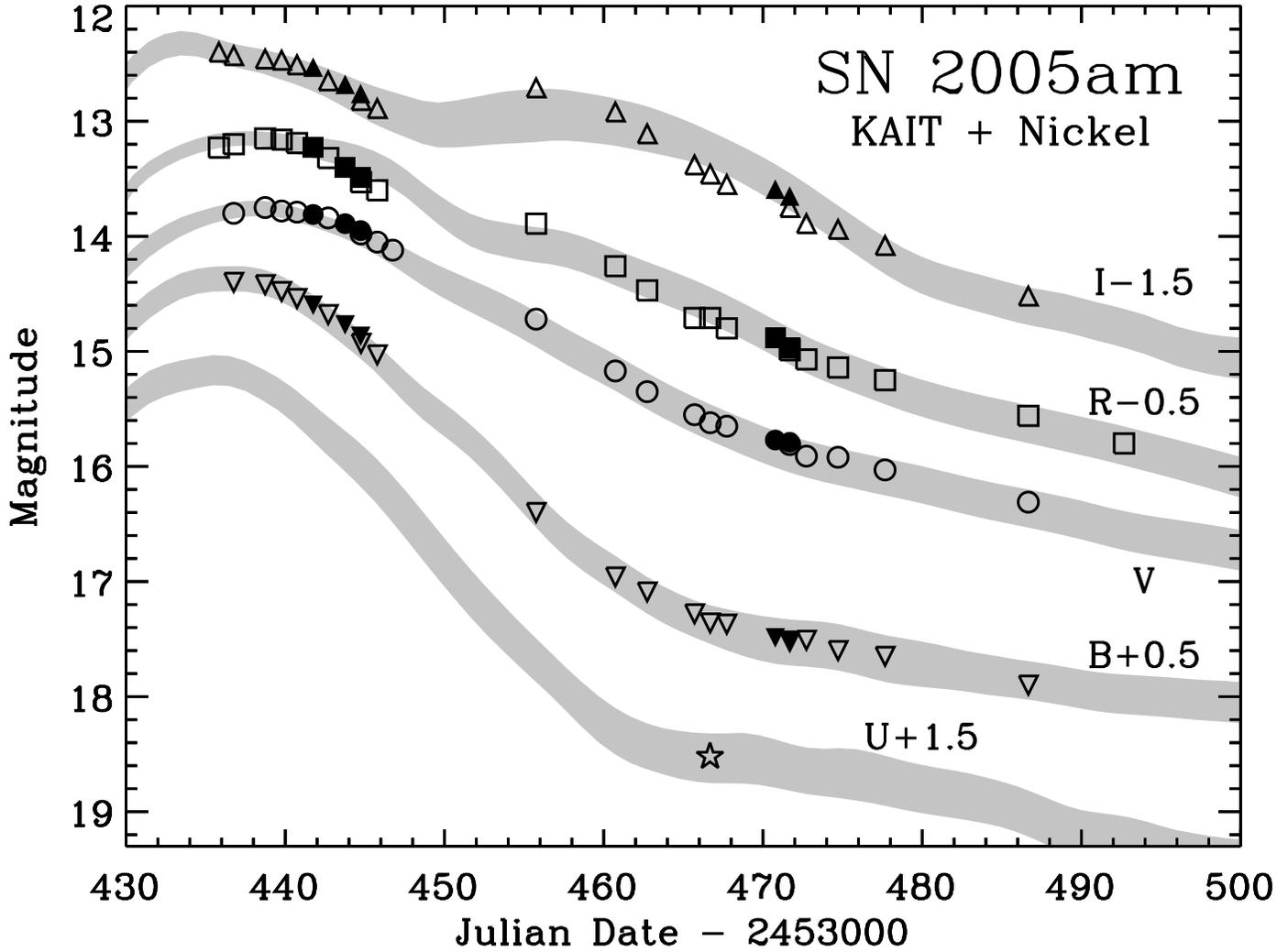}{6.4in}{-270}{80}{80}{520}{10}}
\caption{The ground-based photometry of SN 2005am. The solid
symbols are from the Nickel telescope, and the open symbols are from KAIT. 
Also overplotted are the MLCS fits.}
\end{figure}

\newpage
\begin{figure}
{\plotfiddle{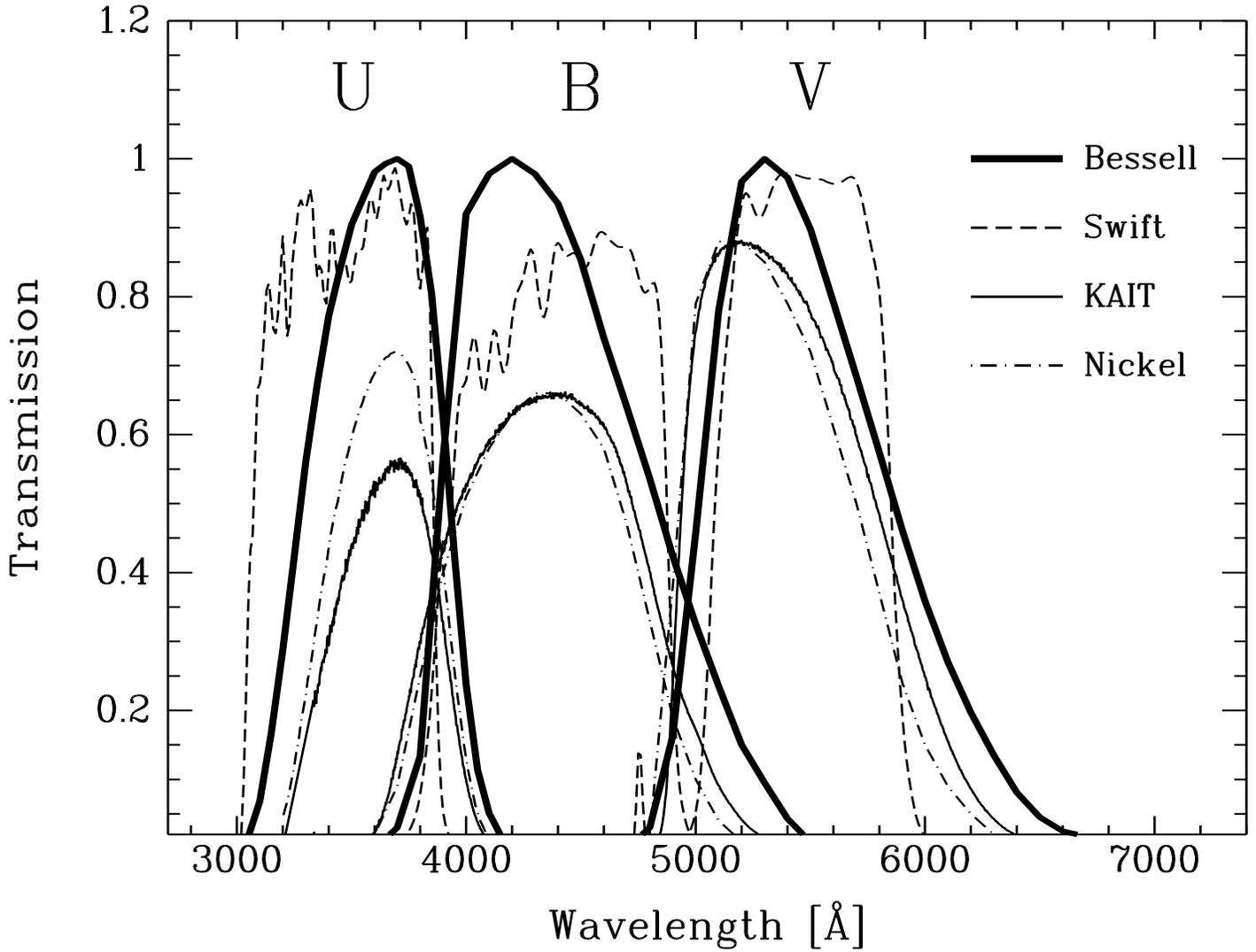}{6.4in}{-90}{80}{80}{-70}{510}}
\caption{The $U$, $B$, and $V$ filter transmission curves. 
Plotted are the filters used by KAIT (thin solid lines),
Nickel (dash-dotted lines), and UVOT (dashed lines).
The Bessell (1990) descriptions are plotted as thick solid
lines.}
\end{figure}

\clearpage
\newpage
\begin{figure}
\caption{Sample PSFs of stars in the
UVOT obs1 $V$ image. (1) A 
very bright star with ghost wings (a), ghost ring (b), and ring (c);
(2) a pair of bright stars with ghost rings (b) and rings (c); 
(3) a bright star with ring (c); and (4) stars with no ghost 
emission, but with varying profiles.}
\end{figure}

\clearpage
\newpage
\begin{figure}
{\plotfiddle{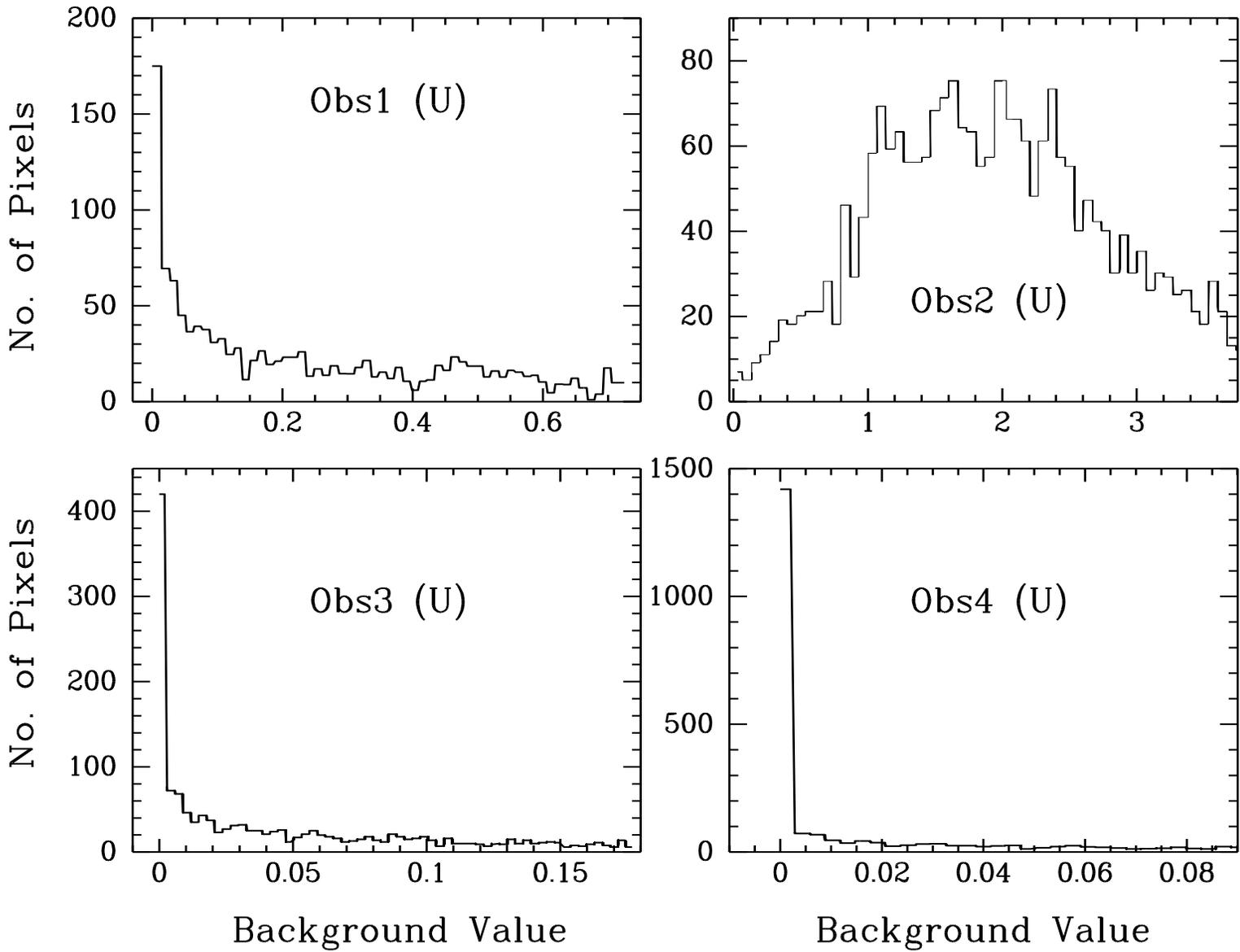}{6.4in}{-90}{80}{80}{-70}{510}}
\caption{The histogram of sky background around star \#5 in the 
four $U$-band observations. Only obs2 has a Gaussian-like 
distribution. The other distributions peak and truncate at
a zero sky value. }
\end{figure}

\clearpage
\newpage
\begin{figure}
{\plotfiddle{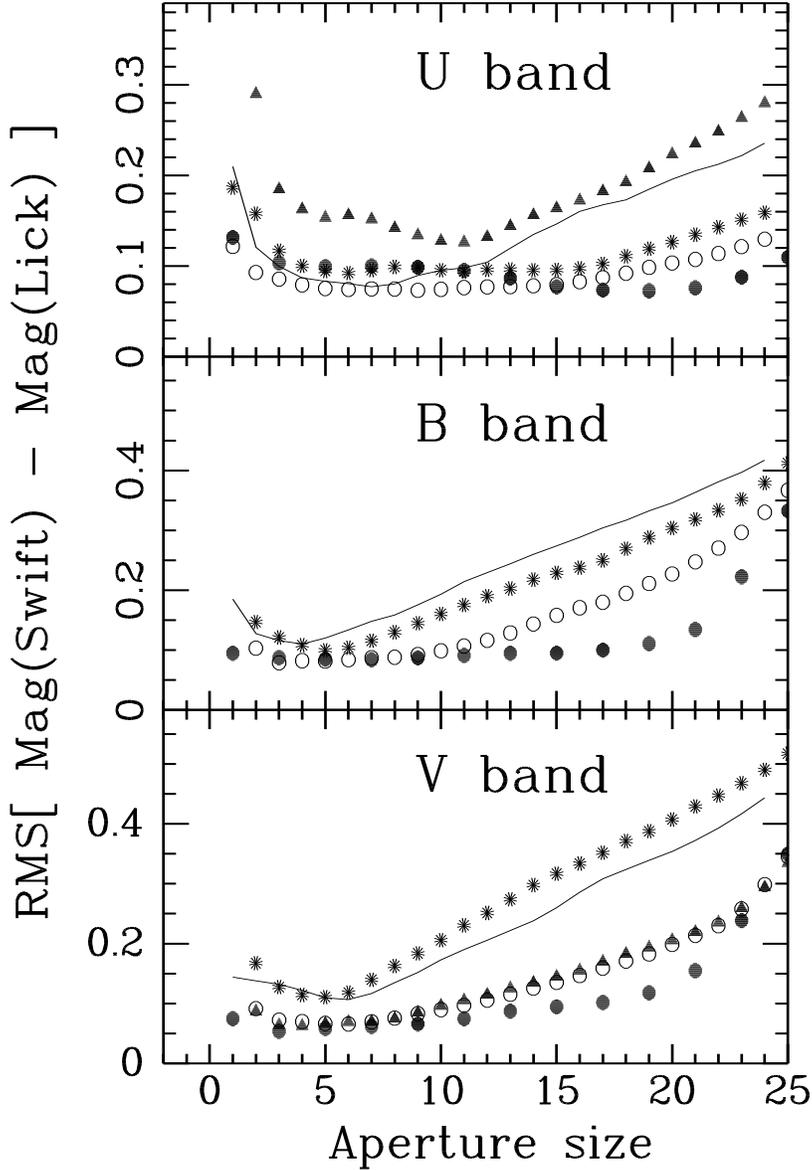}{6.4in}{-90}{80}{80}{-70}{510}}
\caption{The RMS of the differences between the UVOT photometry and
the Lick calibrations, as a function of aperture size. Obs1 is shown
as open circles, obs2 as solid circles, obs3 as stars, obs4
as solid triangles, and obs5 as solid lines. For obs2, the apertures are displayed as 
APT(used)$\times$2 $-$ 1. The smallest RMS is achieved when an
aperture radius of 5 pixels (3 pixels for 2$\times$2 binned data)
is used to do photometry for the $B$ and $V$ bands. The $U$-band
RMS has a flat distribution for aperture sizes in the 5--11 pixel 
range. }
\end{figure}

\clearpage
\newpage
\begin{figure}
{\plotfiddle{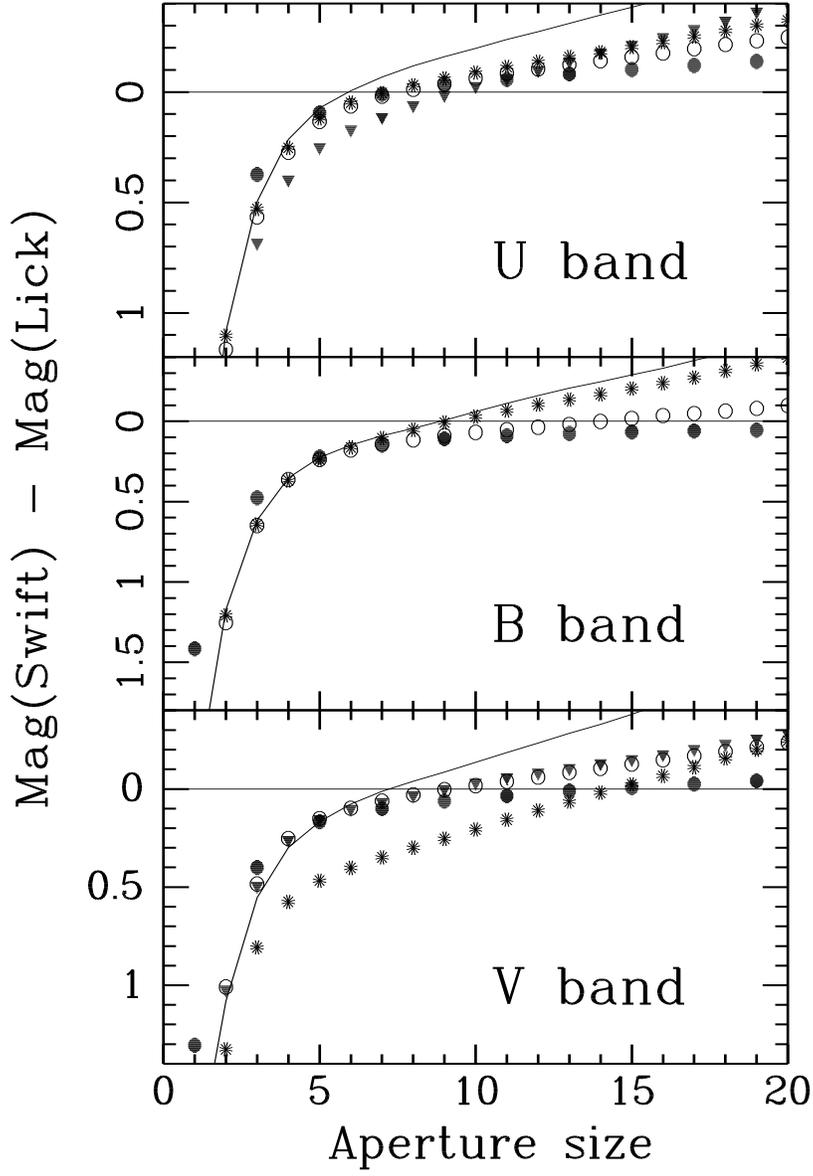}{6.4in}{-90}{80}{80}{-70}{510}}
\caption{The difference between the UVOT photometry and the 
Lick calibrations, as a function of aperture size.  The same 
symbols have been used as in Figure 6.
}
\end{figure}

\clearpage
\newpage
\begin{figure}
{\plotfiddle{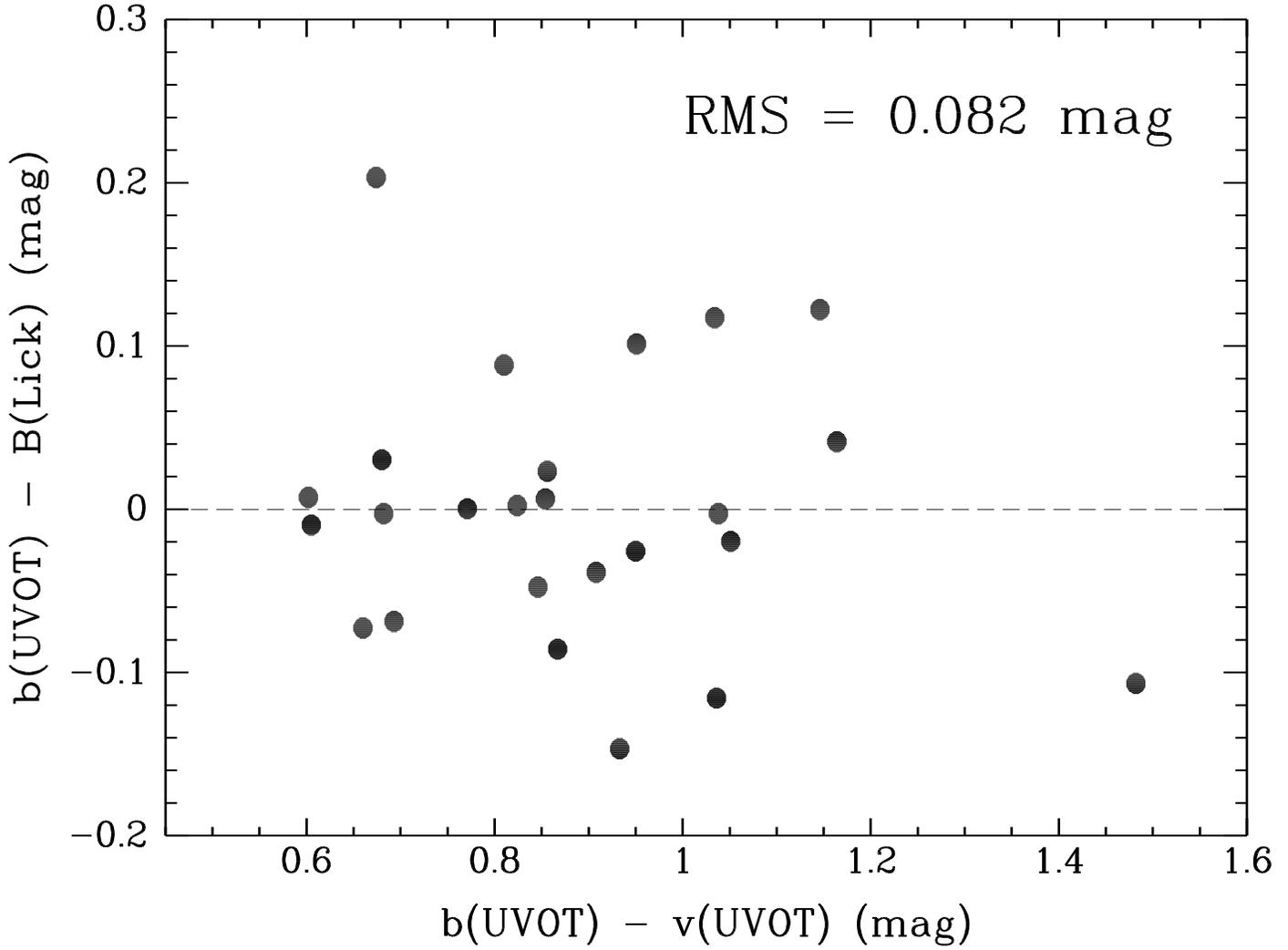}{6.4in}{-90}{80}{80}{-70}{510}}
\caption{The residual $b(UVOT) - B(Lick)$ versus the 
$b(UVOT) - v(UVOT)$ color in obs1 $B$. The RMS is 0.082 mag, and
there is no apparent correlation between the residuals and
the colors.}
\end{figure}

\clearpage
\newpage
\begin{figure}
{\plotfiddle{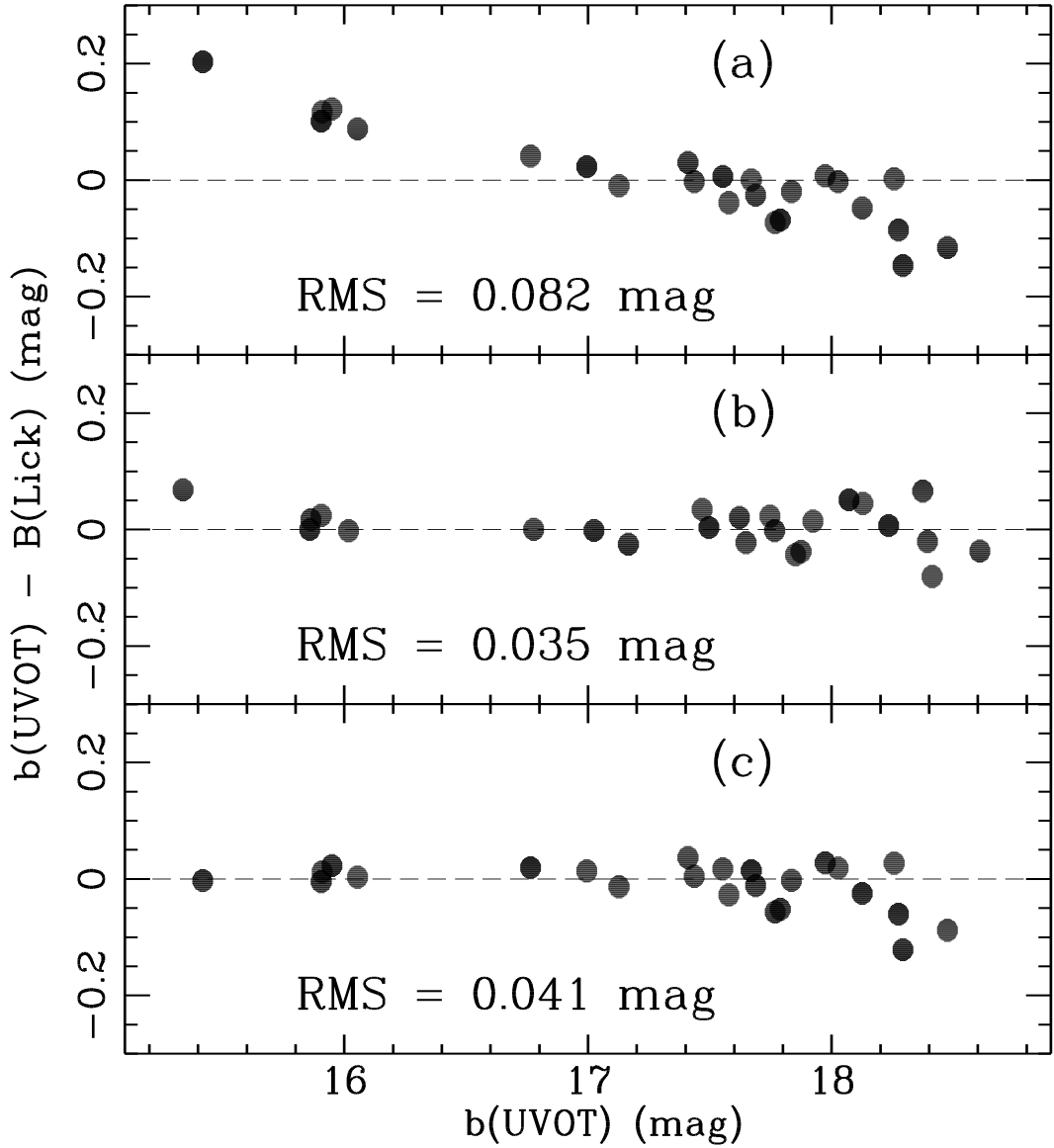}{6.4in}{-90}{80}{80}{-70}{510}}
\caption{Panel (a): the residual $b(UVOT) - B(Lick)$ versus the 
magnitude $b(UVOT)$ in obs1 $B$, which shows a strong correlation. Panel (b):
After a linear correlation is removed, the RMS is significantly improved.
Panel (c): the residuals after coincidence-loss correction. No apparent correlation
is present.}
\end{figure}

\clearpage
\newpage
\begin{figure}
\caption{Finder charts for the fields near SN 2005cf. Both images 
were taken in the R band with the Nickel telescope on 2005 July 11 (field of view
$6\arcmin.3\times6\arcmin.3$). The left panel includes SN 2005cf,
while the right panel does not. Both panels are within the field 
of view for UVOT. North is up and east is to the left.
SN 2005cf and the local standard stars listed in Table 7 are labeled. }
\end{figure}

\clearpage

\newpage
\begin{figure}
{\plotfiddle{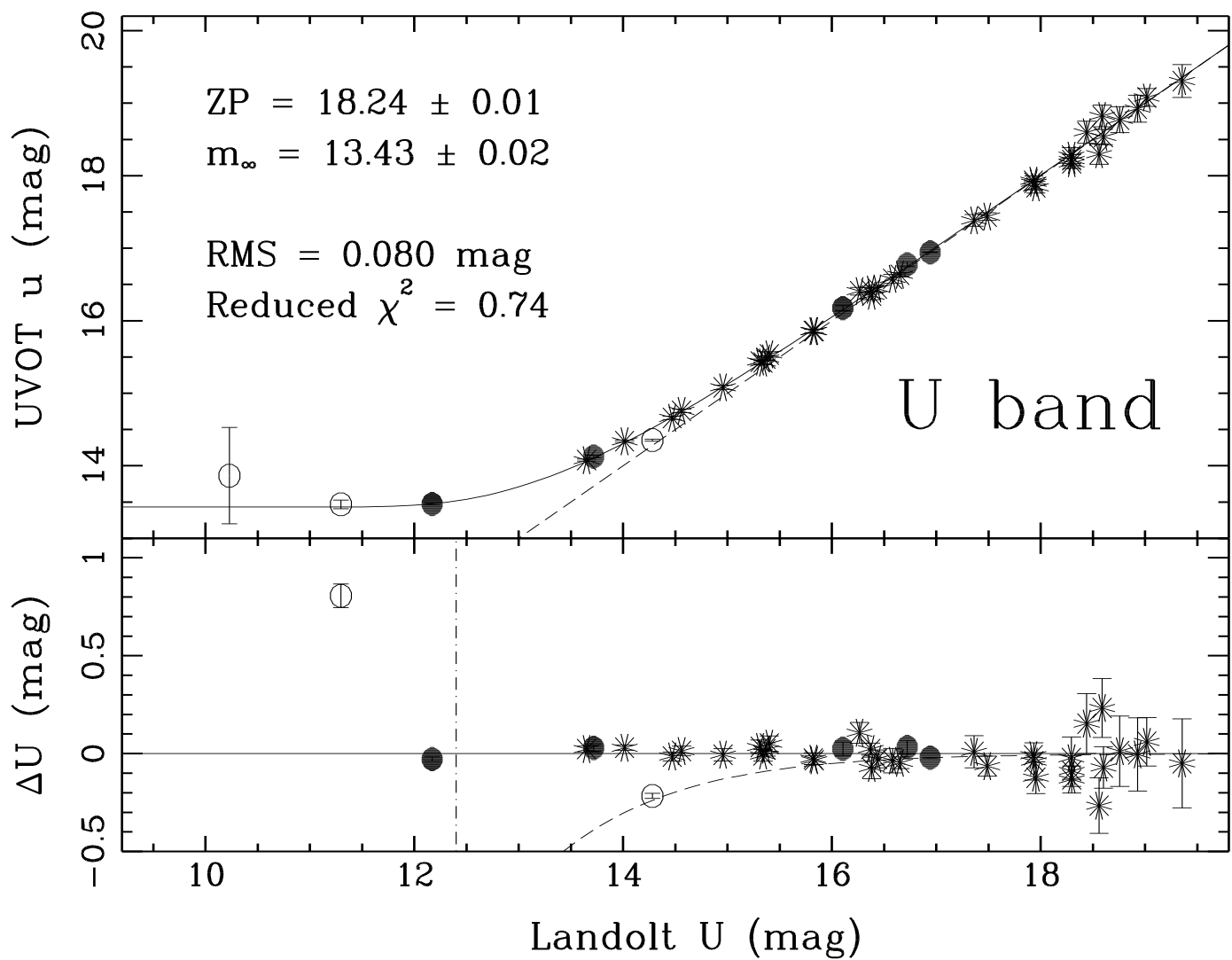}{6.4in}{-90}{80}{80}{-70}{510}}
\caption{The coincidence-loss correction for the $U$ band. The upper 
panel shows the Landolt $U$ versus the UVOT $u$ magnitudes for
the Landolt standard stars (open circles) and the SN 2005cf field
stars (stars). The solid line shows our model coincidence-loss
correction, while the dashed line shows Landolt $U$ = UVOT $u$. 
The open circles are the Landolt standard stars that are not used
in fitting the coincidence-loss correction due to their extreme
brightness or color. The lower panel shows the residuals of the fit.
The dash-dotted line marks the Landolt $U$-band magnitude
whose corresponding UVOT $u$-band magnitude is 0.1 mag fainter
than m$_\infty$. }

\end{figure}

\clearpage

\newpage
\begin{figure}
{\plotfiddle{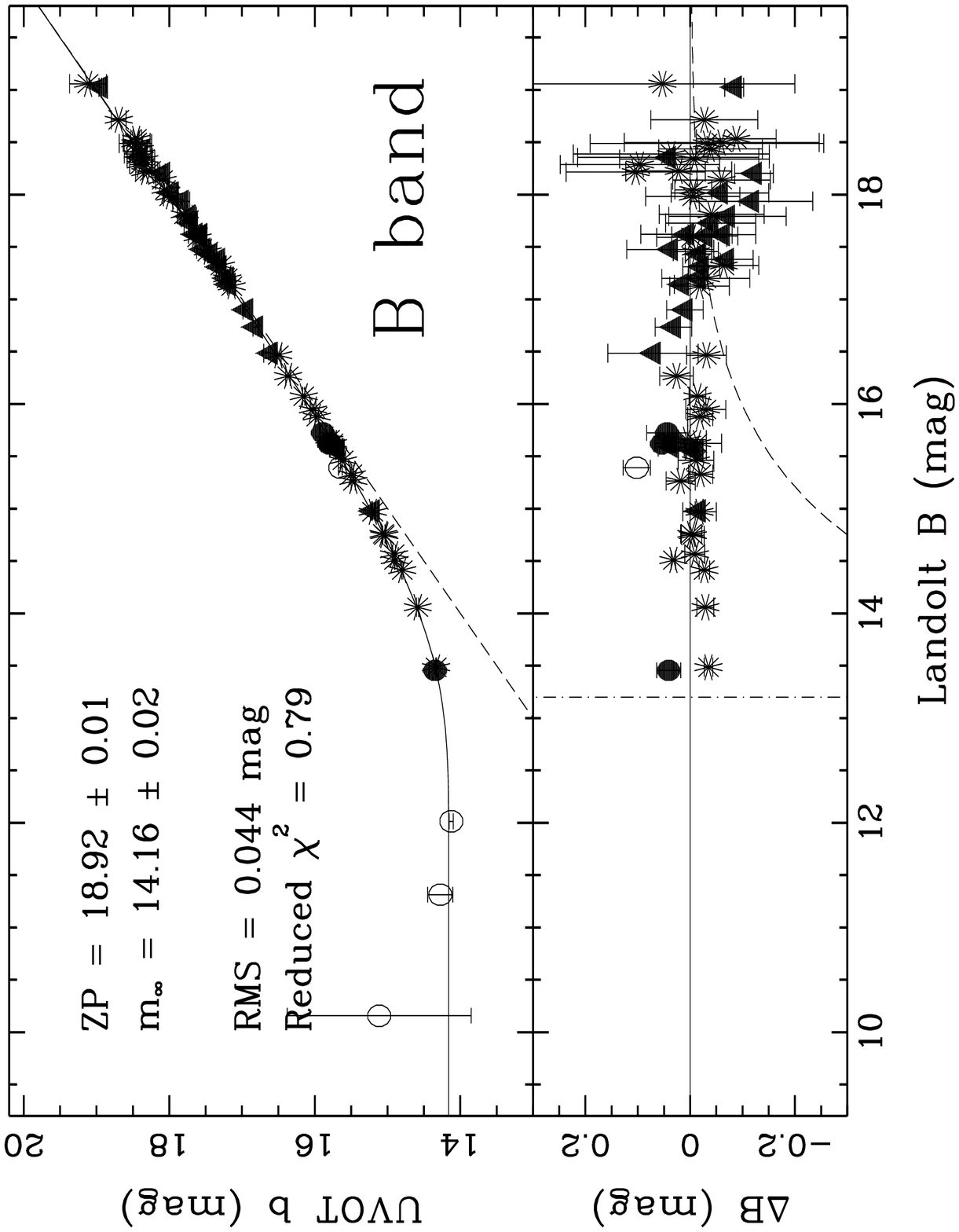}{6.4in}{-90}{80}{80}{-70}{510}}
\caption{Same as Figure 11 but for the $B$ band. The triangles are for the 
SN 2005am field stars. }
\end{figure}

\newpage
\begin{figure}
{\plotfiddle{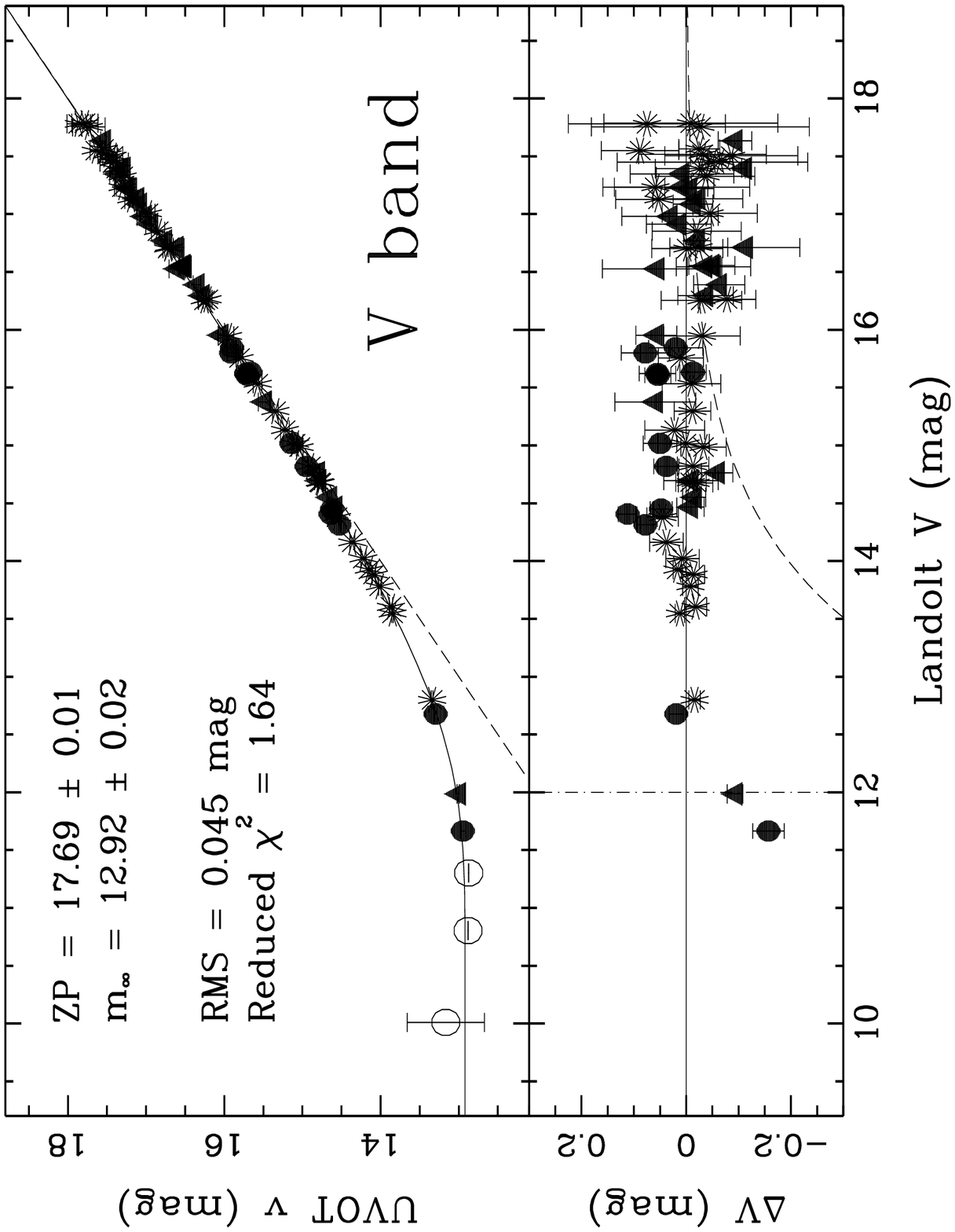}{6.4in}{-90}{80}{80}{-70}{510}}
\caption{Same as Figure 11 but for the $V$ band. The triangles are for the
SN 2005am field stars.} 
\end{figure}

\newpage
\begin{figure}
{\plotfiddle{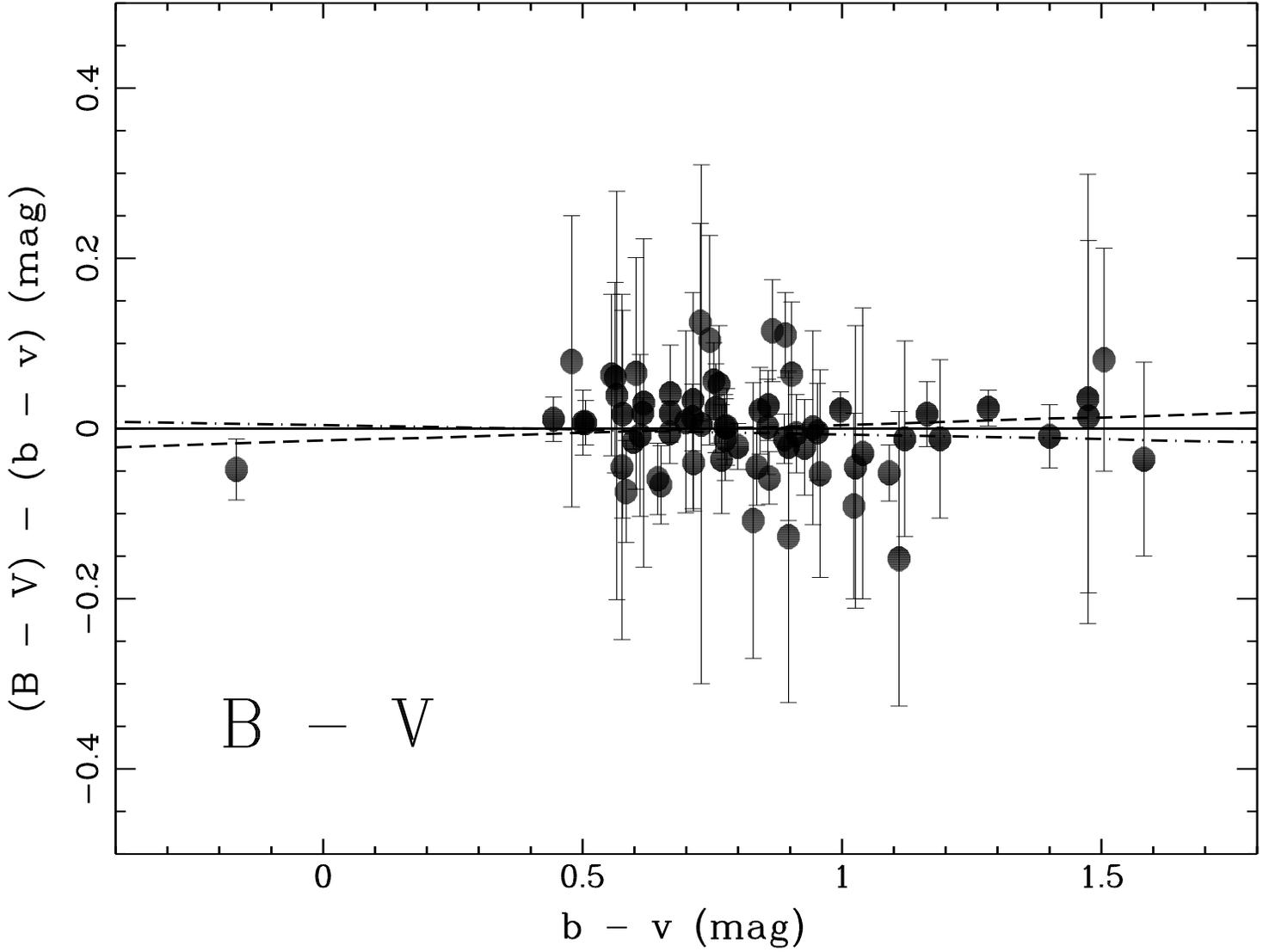}{6.4in}{-90}{80}{80}{-70}{510}}
\caption{The difference between the $(B - V)$ and
$(b - v)$ color, as a function of $(b - v)$. Also overplotted
are three fitting functions: the solid line is 
$(B - V)$ = $(b - v)$, the dash-dotted line is
$(B - V)$ = 0.004 + 0.989$(b - v)$, while the dashed line is
$(B - V)$ = $-$0.014 + 1.0184$(b - v)$. }
\end{figure}

\newpage
\begin{figure}
{\plotfiddle{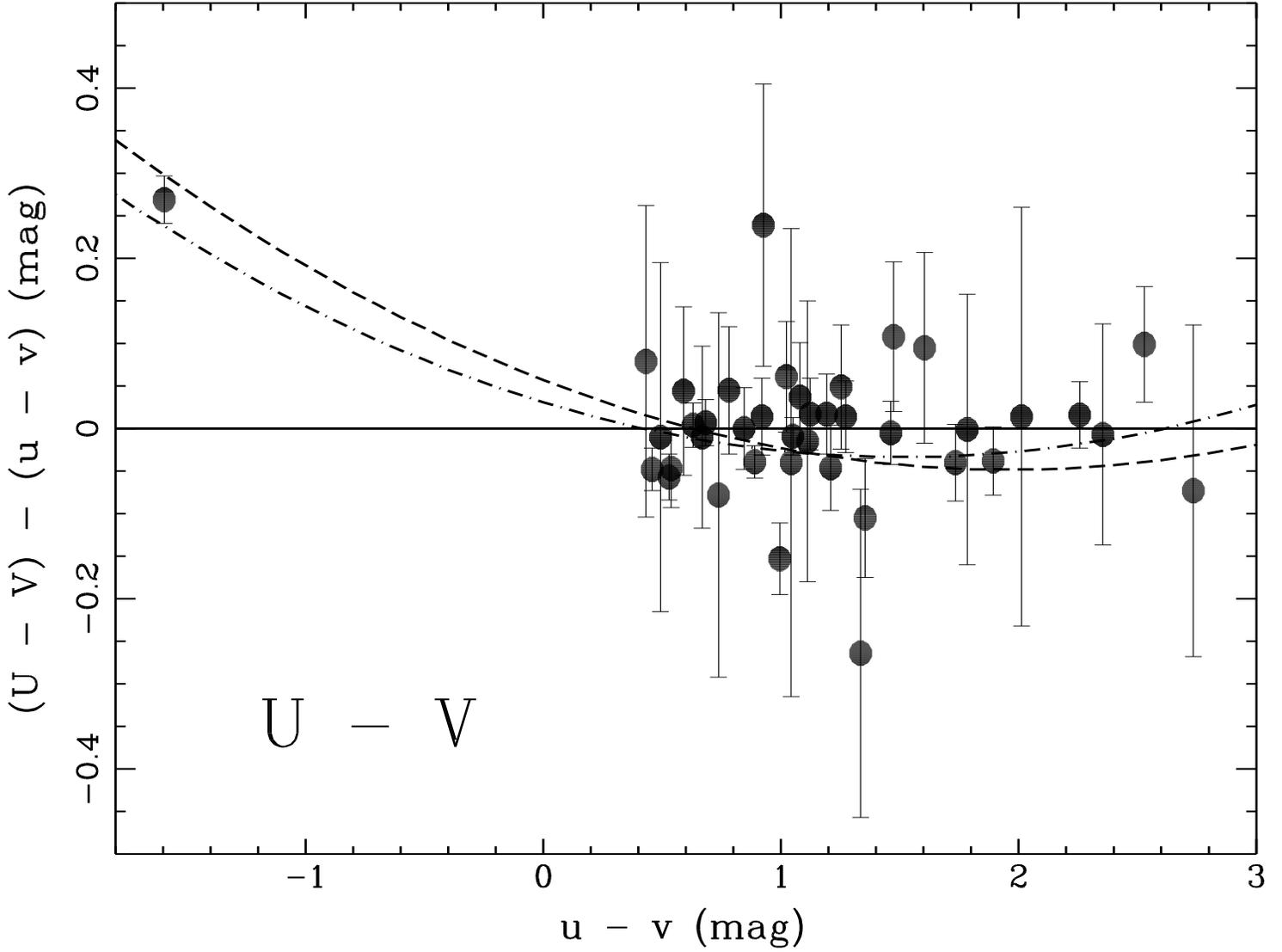}{6.4in}{-90}{80}{80}{-70}{510}}
\caption{The difference between the $(U - V)$ and 
$(u - v)$ color, as a function of $(u - v)$. Also overplotted
are  three fitting functions: the solid line is 
$(U - V)$ = $(u - v)$, the dash-dotted line is
$(U - V)$ = 0.031 + 0.9150$(u - v)$ + 0.028$(u - v)^2$,
while the dashed line is
$(U - V)$ = 0.057 + 0.8926$(u - v)$ + 0.0274$(u - v)^2$.}
\end{figure}

\newpage
\begin{figure}
{\plotfiddle{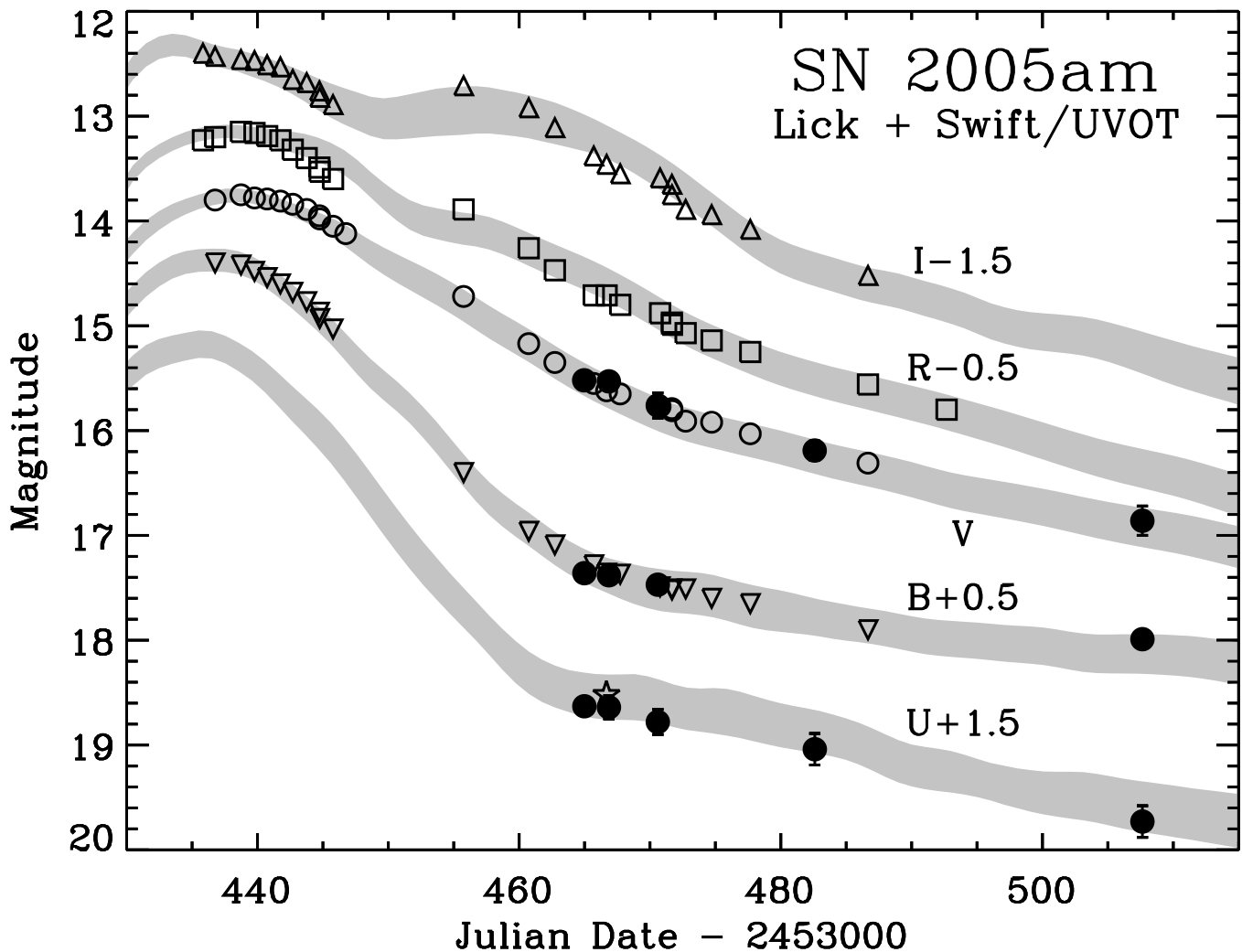}{6.4in}{-270}{80}{80}{520}{10}}
\caption{The ground-based (open symbols) and the UVOT
(solid circles) photometry for SN 2005am. Also overplotted
are the MLCS fits. }
\end{figure}

\newpage
\begin{figure}
\caption{A finder chart (6$\arcmin\times$6$\arcmin$) 
for the field of GRB 050603. This comes from
a 2055.79~s UVOT $V$-band image that started at 01:25:45 
on 2005 June 4. North is up and east is to the left.
The optical afterglow (OA) and star S1 are labeled.
}
\end{figure}

\newpage
\begin{figure}
{\plotfiddle{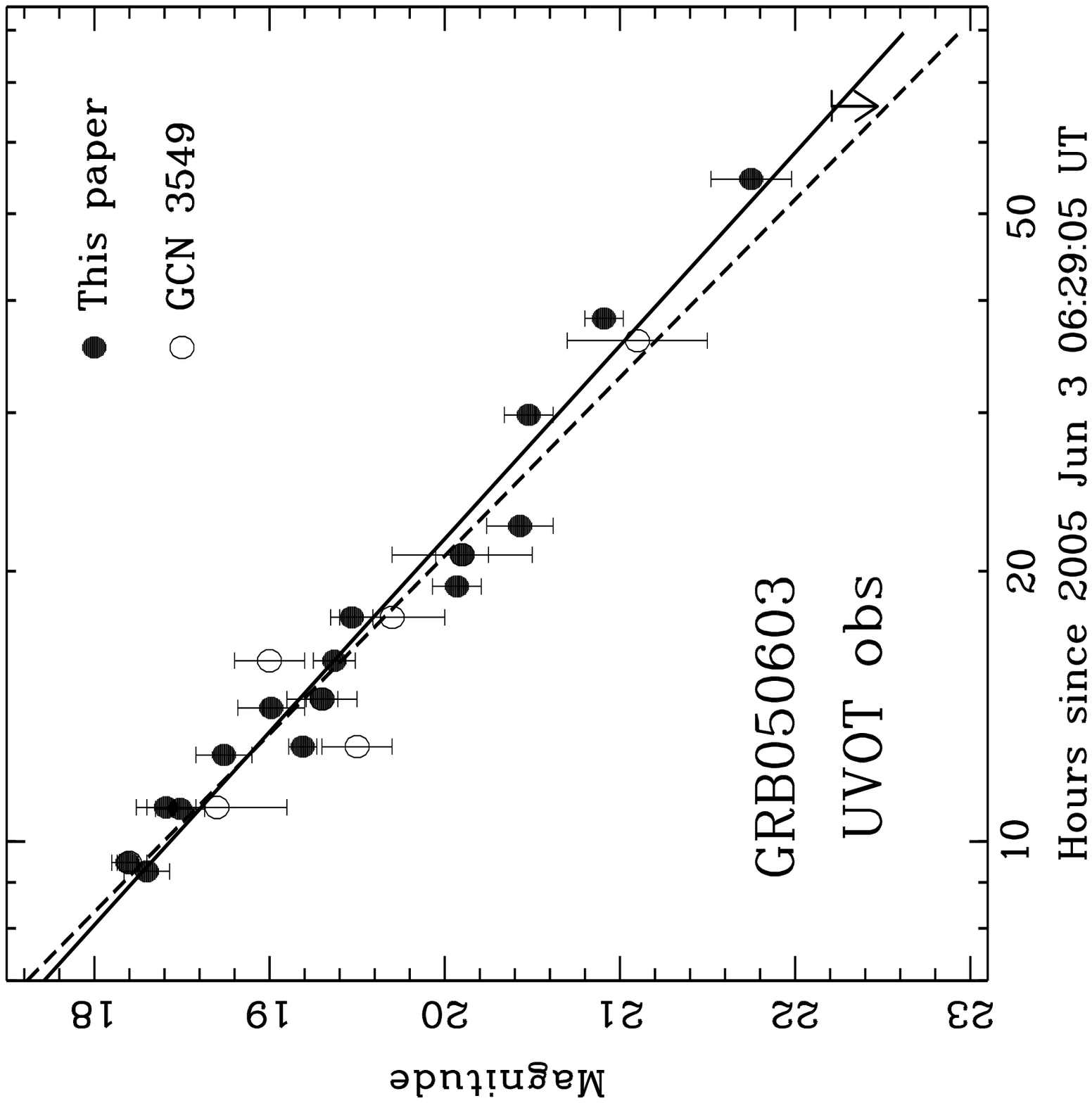}{6.4in}{-90}{80}{80}{-70}{510}}
\caption{The UVOT $V$-band photometry for GRB 050603. The solid
circles are from photometry reported in this paper, while the open circles
are from GCN 3549 (Brown et al. 2005c). The solid line is a
power-law fit to the solid circles, with $\alpha$ = $-$1.86$\pm$0.06,
while the dashed line is a power-law fit to the open circles,
with $\alpha$ = $-$2.01$\pm$0.22.}
\end{figure}


\begin{thebibliography}{}

\bibitem{}Barnard, V., et al., 2005, GCN Circ. 3515
\bibitem{}Barthelmy, S. D., et al., 2005, astro-ph/0507410
\bibitem{}Bertin, E., \& Arnouts, S 1996, A\&AS, 117, 393
\bibitem{}Bessell, M. S. 1990, \pasp, 102, 1181
\bibitem{}Berger, E. 2005, GCN Circ. 3517
\bibitem{}Berger, E., \& Becker, G. 2005, GCN Circ. 3520
\bibitem{}Berger, E., \& McWilliam, A. 2005, GCN Circ. 3511
\bibitem{}Brown, P. J., et al., 2005a, astro-ph/0506029
\bibitem{}Brown, P. J., et al., 2005b, GCN Circ. 3516
\bibitem{}Brown, P. J., et al., 2005c, GCN Circ. 3549
\bibitem{}Burrows, D. N., et al., 2005, astro-ph/0508071
\bibitem{}Cameron, P. B. 2005, GCN Circ. 3513
\bibitem{}Filippenko, A. V. 2003, in {\it From Twilight
  to Highlight: The Physics of Supernovae}, ed. W. Hillebrandt and B.
   Leibundgut (Berlin: Springer-Verlag), 171
\bibitem{}Filippenko, A. V. 2005, in {\it The Fate of the Most Massive
    Stars}, ed. R. Humphreys and K. Stanek (San Francisco: ASP), 34
\bibitem[Filippenko et al.~2001]{fil01}Filippenko, A. V., Li, W., Treffers,
R. R., \& Modjaz, M. 2001, in Small-Telescope Astronomy on Global Scales,
ed. W.-P. Chen, C. Lemme, \& B. Paczy\'{n}ski (San
Francisco: ASP), 121 
\bibitem{}Filippenko, A. V., et al. 1992, AJ, 104, 1543
\bibitem{}Gotz, D., \&  Mereghetti, S. 2005, GCN Circ. 3510
\bibitem{}Jha, S. 2002, PhD thesis, Harvard University
\bibitem{}Jha, S., Riess, A. G., \& Kirshner, R. P. 2006, submitted
\bibitem{}Jha, S., et al., 2005, AJ, in press
\bibitem{}Landolt, A. U. 1992, \aj, 104, 340
\bibitem{}Lazzati, D., Rossi, E., Covino, S. Ghisellini, G., \& Malesani, D. 2002, A\&A, 396, L5
\bibitem{}Li, W., Filippenko, A. V., Chornock, R., \& Jha, S. 2003b, \apj, 586, L9
\bibitem[Li et al.~2000]{wli00} Li, W., et al. 2000, in Cosmic Explosions,
ed. S.~S. Holt \& W.~W.~Zhang (New York: American Institute of Physics), 103 
\bibitem{}Li, W., et al., 2003a, \pasp, 115, 453
\bibitem{}Matheson, T., et al., 2005, \apj, 599, 394
\bibitem{}Modjaz, M., Kirshner, R., \& Challis, P. 2005a, IAU Circ. 8491
\bibitem{}Modjaz, M., Kirshner, R., Challis, P., \& Berlind, P. 2005b, IAU Circ. 8534
\bibitem{}Martin, R., Yamaoka, H., \& Itagaki, K. 2005, IAU Circ. 8490
\bibitem{}Mason, K. O., et al., 2001, A\&A, 365, L36
\bibitem{}Pugh, H., \& Li, W. 2005, IAU Circ. 8534
\bibitem{}Retter, A., et al. 2005, GCN Circ. 3509
\bibitem{}Roming, P. W. A., et al., 2005, astro-ph/0507413
\bibitem{}Stetson, P. B. 1987, \pasp, 99, 191

\end{thebibliography}
\end{document}